\documentclass[fleqn]{mnras}

\usepackage{graphicx}
\usepackage{newtxtext}
\usepackage[cmintegrals]{newtxmath}

\usepackage{bbm}

\usepackage{esint}
\usepackage{verbatim}
\usepackage{bm}

\usepackage{threeparttable}
\usepackage{tabularx}
\usepackage{booktabs}
\usepackage{caption}

\usepackage{scalerel}

\usepackage{booktabs}

\makeatletter
\DeclareRobustCommand*{\bfseries}{%
  \not@math@alphabet\bfseries\mathbf
  \fontseries\bfdefault\selectfont
  \boldmath
}
\makeatother

\catcode`\@=11
\newcommand\gsim{\ifmmode{\mathrel{\mathpalette\@versim>}}
    \else{$\mathrel{\mathpalette\@versim>}$}\fi}
\newcommand\lsim{\ifmmode{\mathrel{\mathpalette\@versim<}}
    \else{$\mathrel{\mathpalette\@versim<}$}\fi}
\catcode`\@=12

\DeclareMathAlphabet{\mathpzc}{OT1}{pzc}{m}{it}

\DeclareMathAlphabet{\mathcal}{OMS}{cmsy}{m}{n}% per il numero di Mach

\newcommand\e{{\rm e}}

\newcommand\M{\mathcal{M}}
\newcommand\gammaAD{\gamma_{\rm ad}}

\newcommand\Ledd{L_{\rm Edd}}
\newcommand\mugas{\langle\mu\rangle}
\newcommand\mH{m_{\rm p}}
\newcommand\sigmaThom{\sigma_{\rm T}}

\newcommand\cp{{\cal C}_p}
\newcommand\cV{{\cal C}_V}

\newcommand\kB{k_{\rm B}}
\newcommand\pinf{p_\infty}
\newcommand\rhoinf{\rho_\infty}
\newcommand\rhotil{\tilde{\rho}}
\newcommand\cs{c_{\rm s}}
\newcommand\cinf{c_\infty}
\newcommand\MBp{\dot{M}_{\rm B}}
\newcommand\rB{r_{\rm B}}
\newcommand\Rinf{R_{\rm inf}}

\newcommand\Tinf{T_\infty}
\newcommand\TV{T_{\rm V}}
\newcommand\sigV{\sigma_{\rm V}}
\newcommand\tWsg{\widetilde{W}_{\rm *g}}

\newcommand\fmin{f_{\rm min}}
\newcommand\gmin{g_{\rm min}}
\newcommand\xmin{x_{\rm min}}
\newcommand\Mmin{{\cal M}_{\rm min}}
\newcommand\rmin{r_{\rm min}}
\newcommand\lcr{\lambdaup_{\hspace{0.3mm}{\rm cr}}}
\newcommand\Lcr{\Lambda_{\rm cr}}
\newcommand\lt{\lambdaup_{\hspace{0.2mm}{\rm t}}}

\newcommand\rhos{\rho_*}

\newcommand\rhoNFW{\rho_{\rm NFW}}
\newcommand\xiNFW{\xi_{\rm NFW}}
\newcommand\RNFW{{\cal R}_{\rm NFW}}
\newcommand\MNFW{M_{\rm NFW}}
\newcommand\rNFW{r_{\rm NFW}}
\newcommand\rt{r_{\rm t}}

\newcommand\rhog{\rho_{\rm g}}
\newcommand\xig{\xi_{\rm g}}
\newcommand\Rg{{\mathcal R}_{\rm g}}

\newcommand\rhoDM{\rho_{\rm DM}}
\newcommand\rhon{\rho_{\rm n}}

\newcommand\rs{r_*}
\newcommand\rg{r_{\rm g}}
\newcommand\Reff{R_{\rm e}}

\newcommand\PsiT{\Psi_{\rm T}}

\newcommand\Psig{\Psi_{\rm g}}
\newcommand\Psin{\Psi_{\rm n}}

\newcommand\Ms{M_*}
\newcommand\MDM{M_{\rm DM}}
\newcommand\MBH{M_{\rm {BH}}}
\newcommand\Mg{M_{\rm g}}
\newcommand\MT{M_{\rm T}}

\newcommand\MR{{\cal R}}
\newcommand\Rm{{\cal R}_{\rm m}}

\newcommand\sigBH{\sigma_{\rm BH}}
\newcommand\sigg{\sigma_{\rm g}}
\newcommand\sigr{\sigma_{\rm r}}

\newcommand\Ks{K_*}

\newcommand\Wsg{W_{\rm *g}}
\newcommand\WsBH{W_{\rm *BH}}

\newcommand\sigp{\sigma_{\rm p}}
\newcommand\sigpBH{\sigma_{\rm pBH}}
\newcommand\sigpg{\sigma_{\rm pg}}

\newcommand\Hf{{\cal H}}

\newcommand\Wz{W_0}
\newcommand\Wmu{W_{-1}}

\newcommand\Fg{{\cal F}_{\rm g}}
\newcommand\bc{\beta_{\rm c}}

\newcommand\s{\mathcal{S}}
\newcommand\sinf{{\mathcal S}_\infty}
\newcommand\E{\mathcal{E}}

\newcommand\Qn{{\cal Q}_{\rm n}}

\newcommand\er{{\bm e}_r}
\newcommand\Grad{\boldsymbol{\nabla}} %\vv\hspace{0.15mm}\cdot\Grad

\newcommand\Qdot{\mathcal{Q}}

\newcommand\Og{{\cal O}}

\newcommand\Mest{M_{\rm est}}

\newcommand\tT{\tilde{T}}

\title[Bondi accretion in galaxies with a massive BH]{On the polytropic Bondi accretion in two-component galaxy models with a central massive BH}

\author[A. Mancino, L. Ciotti \& S. Pellegrini]{Antonio Mancino$^{1,2}$, Luca Ciotti$^1$ \& Silvia Pellegrini$^{1,2}$ 
\\
$^1$Department of Physics and Astronomy, University of Bologna, via Gobetti 93/3, 40129 Bologna, Italy
\\
$^2$Istituto Nazionale di Astrofisica (INAF), Osservatorio di Astrofisica e Scienza dello Spazio di Bologna (OAS), Via Gobetti 93/3, Bologna 40129, Italy}

\date{Accepted 2022 March 3}

\begin{document}
\maketitle
\label{firstpage}

%%%%%%%%%%%%%%%%%%%%%%%%%%%%%%%%%%%%%%%%%%%%%%%%%%%%%%%%%%%%%%%%%%%%%%%%%%%%%%%%%%%%%%%%%%%%%%
%*******************************************SOMMARIO******************************************
%%%%%%%%%%%%%%%%%%%%%%%%%%%%%%%%%%%%%%%%%%%%%%%%%%%%%%%%%%%%%%%%%%%%%%%%%%%%%%%%%%%%%%%%%%%%%%

  \begin{abstract}

  \noindent
  In many investigations involving accretion on a central point mass, 
  ranging from observational studies to cosmological simulations, including semi-analytical 
  modelling, the classical Bondi accretion theory is the standard tool widely adopted.
  Previous works generalised the theory to include the 
  effects of the gravitational field of the galaxy hosting a central black hole, and of electron scattering
  in the optically thin limit.
  Here we apply this extended Bondi problem, in the general polytropic case, to a class of new
  two-component galaxy models 
  recently presented. In these models, a Jaffe stellar density profile is
  embedded in a dark matter halo such that the total density distribution follows a $r^{-3}$
  profile at large radii; the stellar dynamical quantities can be expressed in a fully analytical way. 
  The hydrodynamical properties of the flow are set by imposing that 
  the gas temperature at infinity is proportional to the virial temperature of the stellar component.  
  The isothermal and adiabatic (monoatomic) cases can be solved analytically, in
  the other cases we explore the accretion solution numerically.
  As non-adiabatic accretion inevitably leads to an exchange of 
  heat with the ambient, we also discuss some important thermodynamical properties of 
  the polytropic Bondi accretion,
  and provide the expressions needed to compute the amount of heat exchanged with the 
  environment, as a function of radius.
  The results can be useful for the subgrid treatment of accretion in numerical simulations, 
  as well as for the interpretation of observational data. 
  \end{abstract}


%%%%%%%%%%%%%%%%%%%%%%%%%%%%%%%%%%%%%%%%%%%%%%%%%%%%%%%%%%%%%%%%%%%
  \section{Introduction}\label{sec:Intro}
%%%%%%%%%%%%%%%%%%%%%%%%%%%%%%%%%%%%%%%%%%%%%%%%%%%%%%%%%%%%%%%%%%%

%%%%%%%%%%%%%%%%%%%%%%%%%%%%%%%%%%%%%%%%%%%%%%%%%%%%%%%%%%%%%%%%%%%%%%%%%%%%%%%%%%%%%%%%%%%%%
\begin{table*}
 \begin{threeparttable}
\centering
    \caption{Main properties of accretion solutions in one- and two-component
             galaxy models with central MBH.}
    \begin{tabular}{@{}l@{\hskip 4em}c@{\hskip 4em}c@{\hskip 4em}c@{\hskip 4em}c}
        \toprule[1.25pt]\midrule[0.3pt]
        & KCP16 & CP17 & CP18 & MCP21 (this paper)\\
        \midrule
        Galaxy models  & Hernquist (1990) & Hernquist (1990), Jaffe (1983) & JJ models (CZ18) & J3 models (CMP19)\\ 
        \midrule                                      
        Type of accretion  & Polytropic & Isothermal & Isothermal & Polytropic \\                                                                                                  
        \midrule
        Number of sonic points & One or two & One or two (Hernquist), One (Jaffe) & One & One or two\tnote{b} \\
 \midrule 
  Sonic radius & Analytic\tnote{a} & Analytic & Analytic & Analytic/numerical\tnote{c} \\
 \midrule 
  $\lt$ & Analytic\tnote{a} & Analytic & Analytic & Analytic/numerical\tnote{c} \\
 \midrule 
  Mach number profile & Numerical & Analytic & Analytic & Analytic/numerical\tnote{c} \\
\bottomrule 
    \end{tabular}
    \vspace{1.5mm}
\begin{tablenotes}[para,flushleft]
      \item[a] The general expression can be written as a function of the polytropic 
                index, but only special cases were given explicitly;\\
      \item[b] Function of the polytropic index $\gamma$;\\
      \item[c] In the isothermal ($\gamma=1$) and monoatomic adiabatic ($\gamma=5/3$) cases 
               it is analytic, in the $1<\gamma<5/3$ case only
               a numerical exploration is possible.
    \end{tablenotes}
  \end{threeparttable}
\label{table:models}
\end{table*}
%%%%%%%%%%%%%%%%%%%%%%%%%%%%%%%%%%%%%%%%%%%%%%%%%%%%%%%%%%%%%%%%%%%%%%%%%%%%%%%%%%%%%%%%%%%%

  Theoretical and observational studies indicate that 
  galaxies host at their centre a massive black hole (MBH) that 
  has grown its mass predominantly through gas accretion (see e.g. 
  Kormendy \& Richstone 1995).  
  A generic accretion flow may be broadly classified as 
  quasi-spherical or axisymmetric, and what mainly determines 
  the deviation from spherical symmetry is the angular momentum of 
  the flow itself. 
  A perfect spherical flow is evidently only possible when the 
  angular momentum is exactly zero. 
  Spherical models are a useful starting 
  point for a more advanced modelling, 
  and thus gas accretion toward a central MBH in galaxies is 
  often modelled with the classical Bondi (1952) solution. 
  For example, in semi-analytical models and cosmological simulations 
  of the co-evolution of galaxies and their central MBHs, the mass 
  supply to the accretion discs is linked to the 
  temperature and density of their environment by making use of the 
  Bondi accretion rate (see e.g. Fabian \& Rees 1995; Volonteri \& Rees 
  2005; Booth \& Schaye 2009; Wyithe \& Loeb 2012; Curtis \& Sijacki 2015; 
  Inayoshi, Haiman \& Ostriker 2016). 
  In fact, in most cases, the resolution of simulations cannot describe in detail 
  the whole complexity of accretion, and so Bondi 
  accretion does represent an important approximation to more 
  realistic treatments (see e.g. Ciotti \& Ostriker 2012; Barai et al. 2012;
  Ram\'{i}rez-Velasquez et al. 2018; Gan et al. 2019 and references therein). 
  Recently, Bondi accretion has been generalised to include the effects on the flow
  of the gravitational field of the host galaxy and of
  electron scattering, at the same time preserving the (relative)
  mathematical tractability of the problem. 
  Such a generalised Bondi problem has been applied to elliptical 
  galaxies by Korol et al. (2016, hereafter KCP16), who 
  discussed the case of a Hernquist (1990) galaxy model, for generic values of the 
  polytropic index. 
  Restricting to isothermal accretion, also taking 
  into account the effects of radiation pressure due to electron scattering,
  Ciotti \& Pellegrini (2017, hereafter CP17) showed that 
  the whole accretion solution can be found analytically for the 
  Jaffe (1983) and Hernquist galaxy models with a central MBH; quite 
  remarkably, not only can the critical accretion parameter be explicitly 
  obtained, but it is also possible to write the radial profile of the Mach 
  number via the Lambert-Euler $W$-\hspace{0.4mm}function (see e.g. Corless et 
  al. 1996). 
  Then, Ciotti \& Pellegrini (2018, hereafter CP18) further extended the 
  isothermal accretion solution to the case of Jaffe's 
  two-component (stars plus dark matter) galaxy models  
  (Ciotti \& Ziaee Lorzad 2018, hereafter CZ18). 
  In these `JJ' models, a Jaffe stellar profile is embedded in a DM halo such that the total density 
  distribution is also a Jaffe profile, and all the relevant 
  dynamical properties can be written with analytical expressions.
  CP18 derived all accretion properties analytically,  
  linking them to the dynamical and structural properties of the host galaxies.
  These previous results are summarised in Table 1.
  
  In this paper we extend the study of CP18 
  to a different family of two-component galaxy models with a central MBH, 
  in the general case of a polytropic gas. 
  In this family (J3 models; Ciotti, Mancino \& Pellegrini 2019, hereafter 
  CMP19) the stellar density follows a Jaffe profile, while the total 
  follows a $r^{-3}$ law at large radii; thus the DM halo (resulting from
  the difference between the total and the stellar distributions) can reproduce the
  Navarro-Frenk-White profile (Navarro, Frenk \& White 1997, hereafter NFW) 
  at all radii.  
  As we are concerned with polytropic accretion, we also clarify some thermodynamical 
  aspect of the problem, not always stressed. In fact, it is obvious
  that for a polytropic index $\gamma\neq\gammaAD$ (the adiabatic index of the gas,
  with $\gammaAD=\cp/\cV$) the flow is not adiabatic, and heat exchanges with the
  environment are unavoidable. We investigate in detail this point, obtaining the expression
  of the radial profile of the heat exchange (i.e. radiative losses) of the fluid elements
  as they move towards the galaxy centre. 
  Qualitatively, an implicit cooling/heating function is contained
  in the polytropic accretion when $\gamma\neq\gammaAD$.
 
  The paper is organised as follows. 
  In Section \ref{sec:Bondi}, we recall the main properties of the polytropic Bondi 
  solution, and in Section \ref{sec:J3} we list the main properties of the J3 models. 
  In Section \ref{sec:BondiJ3}, we set up and discuss the polytropic Bondi problem in 
  J3 galaxy models, while in Section \ref{sec:Entropy}, we investigate some important 
  thermodynamical properties of accretion. The main results are finally summarised in 
  Section \ref{sec:Conclusions}, while some technical detail is given in the Appendix.

%%%%%%%%%%%%%%%%%%%%%%%%%%%%%%%%%%%%%%%%%%%%%%%%%%%%%%%%%%%%%%%%%%%%%%%
  \section{Bondi accretion in galaxies}\label{sec:Bondi}
%%%%%%%%%%%%%%%%%%%%%%%%%%%%%%%%%%%%%%%%%%%%%%%%%%%%%%%%%%%%%%%%%%%%%%%  
  
  In order to introduce the adopted notation, and for consistency with previous works,
  in this Section we summarise the main properties of Bondi accretion,
  both on a point mass (i.e., a MBH) and on a MBH at the centre of a 
  spherical galaxy. In particular, the flow is spherically symmetric,
  and the gas viscosity and conduction are neglected.
  
  \subsection{The classical Bondi accretion}\label{subsec:Bondi_classic}
  
  In the classical Bondi problem, a spatially infinite distribution of 
  perfect gas is accreting onto an isolated central point mass (a 
  MBH in our case) of mass $\MBH$. The pressure $p$ and density $\rho$ 
  are related by 
  \begin{equation}
         p=\frac{\kB\hspace{0.1mm}\rho\hspace{0.4mm}T}{\mugas\hspace{0.15mm}\mH}
          =\pinf\hspace{-0.15mm}\times\left(\frac{\rho}{\rhoinf}\right)^{\hspace{-0.1mm}\gamma}\!,
  \label{eq:pgas}
  \end{equation}
  where $\kB=1.38\times 10^{-16}\,{\rm erg}\hspace{0.7mm}{\rm K}^{-1}$ 
  is Boltzmann's constant, $\mugas$ is the mean molecular 
  weight, $\mH=1.67\times 10^{-24}\,{\rm g}$ is the mass of the proton, 
  and $\gamma \geq 1$ is the polytropic 
  index\footnote{In principle $\gamma \geq 0$; in this paper we consider 
  $0 \leq \gamma < 1$ as a purely academic interval.}.
  Finally, $\pinf$ and $\rhoinf$ are the gas pressure and density 
  at infinity, and $\cs=\sqrt{\gamma\hspace{0.1mm} p/\rho\hspace{0.7mm}}$
  is the local polytropic speed of sound. 
  As some confusion unfortunately occurs in the literature, it is important to
  recall that in general $\gamma$ is {\it not} the adiabatic index $\gammaAD$ of the
  gas\footnote{$\gammaAD \equiv \cp/\cV$ is the ratio of specific heats at 
  constant pressure and volume; for a perfect gas, it always exceeds unity.}
  (e.g. Clarke \& Carswell 2007).
  
  The equation of continuity reads
  \begin{equation}
         4\upi r^2\hspace{-0.3mm}\rho\hspace{0.4mm}\varv=\MBp\hspace{0.2mm},
  \label{eq:continuity}
  \end{equation}
  where $\varv(r)$ is the modulus of the gas radial velocity, and $\MBp$ is 
  the time-independent accretion rate onto the MBH. Bernoulli's
  equation, by virtue of the boundary conditions at infinity, is
  \begin{equation}
        \frac{\varv^2}{2}\hspace{0.15mm}+\int_{\pinf}^p\frac{dp}{\rho}-\frac{G\MBH}{r}=0.
  \label{eq:Bernoulli}
  \end{equation} 
  Notice that, unless $\gamma=\gammaAD$, the integral at the left hand side 
  is {\it not} the enthalpy change per unit mass (see Section \ref{sec:Entropy}). 
  The natural scale length of the problem is the Bondi radius
  \begin{equation}
        \rB \equiv \frac{G\MBH}{\cinf^2},
  \label{eq:rB}
  \end{equation}
  and, by introducing the dimensionless quantities
  \begin{equation}
        x \equiv \frac{r}{\rB}\hspace{0.2mm},
        \qquad\,\,\,
        \rhotil \hspace{0.3mm}\equiv \frac{\rho}{\rhoinf}\hspace{0.2mm},
        \qquad\,\,\,
        \M \equiv \frac{\varv}{\cs}\hspace{0.2mm},
  \end{equation}    
  where $\M$ is the local Mach number, 
  equations \eqref{eq:continuity} and \eqref{eq:Bernoulli}  
  become respectively (for $\gamma \neq 1$)
  \begin{equation}
        x^2\hspace{-0.15mm}\M\hspace{0.35mm}\rhotil^{\hspace{0.15mm}\frac{\gamma+1}{2}}\hspace{-0.4mm}=\hspace{0.2mm}\lambdaup\hspace{0.5mm},
        \qquad\,\,
        \left(\frac{\M^2}{2}+\frac{1}{\gamma-1}\right)\!\rhotil^{\hspace{0.5mm}\gamma-1}\hspace{-0.2mm}=\frac{1}{x}+\frac{1}{\gamma-1}\hspace{0.2mm},
  \label{eq:cont+Bern_lambda}
  \end{equation}   
  where 
  \begin{equation}
        \lambdaup \equiv \frac{\MBp}{4\upi\rB^2\rhoinf\cinf}\hspace{0.2mm},
  \end{equation}
  is the (dimensionless) accretion parameter. Once $\MBH$, $\rhoinf$, and 
  $\cinf$ are assigned, if $\lambdaup$ is known it is possible to
  determine the accretion rate $\MBp$ and derive the profile $\M(x)$,
  thus solving the Bondi (1952) problem. As well known, $\lambdaup$ cannot assume
  arbitrary values. In fact, by elimination of $\rhotil$ in between equations 
  \eqref{eq:cont+Bern_lambda}, one obtains the identity
  \begin{equation}
        g(\M)=\Lambda\hspace{0.2mm}f(x)\hspace{0.4mm},
        \qquad\,\,\,
        \Lambda \equiv \lambdaup^{-\frac{2(\gamma-1)}{\gamma+1}},
  \label{eq:Bondi_eq_poly}
  \end{equation}
  where 
  \begin{equation}
        \begin{cases}
              \displaystyle{g(\M) \equiv \M^{-\frac{2(\gamma-1)}{\gamma+1}}\!\left(\frac{\M^2}{2}+\frac{1}{\gamma-1}\right)}\hspace{0.4mm},
              \\[12pt]
              \displaystyle{f(x) \equiv x^{\frac{4(\gamma-1)}{\gamma+1}}\!\left(\frac{1}{x}+\frac{1}{\gamma-1}\right)}\hspace{0.2mm}.
        \end{cases}
  \label{eq:B_poly}
  \end{equation}
  Since both $g$ and $f$ have a minimum, the solutions of equation \eqref{eq:Bondi_eq_poly} 
  exist only when $\gmin \leq \Lambda\fmin$, i.e. 
  $\Lambda \geq \Lcr \equiv \gmin/\fmin$. For $\gamma < 5/3$, 
  \begin{equation}
        \begin{cases}
              \displaystyle{\gmin=\frac{\gamma+1}{2(\gamma-1)}}\hspace{0.2mm},
              \hspace{2.304cm}\hspace{2mm} \Mmin=1\hspace{0.1mm},
              \\[12pt]
              \displaystyle{\fmin=\frac{\gamma+1}{4(\gamma-1)}}\hspace{-0.2mm}\left(\frac{4}{5-3\gamma}\right)^{\hspace{-0.4mm}\frac{5-3\gamma}{\gamma+1}}\!,
              \hspace{2mm}\hspace{5.85mm} \xmin=\dfrac{5-3\gamma}{4}\hspace{0.2mm};
        \end{cases}
  \end{equation}
  therefore, from equation \eqref{eq:Bondi_eq_poly}, the classical Bondi problem 
  admits solutions only for
  \begin{equation}
        \lambdaup \leq \lcr \equiv \left(\frac{\fmin}{\gmin}\right)^{\hspace{-0.4mm}\frac{\gamma+1}{2(\gamma-1)}} = \hspace{0.1mm}\frac{1}{4}\hspace{-0.2mm}\left(\frac{2}{5-3\gamma}\right)^{\hspace{-0.3mm}\frac{5-3\gamma}{2(\gamma-1)}}.
  \label{eq:lambda_cr}
  \end{equation}
  Notice that for $\gamma=5/3$, $\fmin \to 1$, $\xmin \to 0$, and  
  $\lambdaup \leq 1/4$. When $\gamma>5/3$, instead, $\xmin \to 0$ and
  $\fmin=0$, and so no accretion can take place:  
  $\gamma=5/3$ is then a {\it hydrodynamical limit} for the classical Bondi problem.
   
  For $\lambdaup=\lcr$ (the critical solutions), $\xmin$ indicates 
  the position of the sonic point, i.e. $\M(\xmin)=1$. When $\lambdaup<\lcr$, 
  instead, two regular subcritical solutions exist, one everywhere supersonic
  and another everywhere subsonic; the position $\xmin$ marks the minimum
  and maximum value of $\M$, respectively for these two solutions
  (see e.g. Bondi 1952; Frank, King \& Raine 1992; Krolik 1998).
  
  In the $\gamma=1$ (isothermal) case, $p=\cinf^2\rho$, and $\cs=\cinf$,
  while equation \eqref{eq:Bondi_eq_poly} becomes
  \begin{equation}
        g(\M)=f(x)-\Lambda\hspace{0.4mm}, 
        \qquad\,\, 
        \Lambda\equiv\ln\lambdaup\hspace{0.5mm},
  \label{eq:Bondi_eq_iso}
  \end{equation}
  where now
  \begin{equation}
        \begin{cases}
              g(\M) \equiv \dfrac{\M^2}{2}-\ln\M\hspace{0.2mm},
              \\[10pt]
              f(x) \equiv \dfrac{1}{x}+2\ln x\hspace{0.1mm}.
        \end{cases}
  \label{eq:B_iso}
  \end{equation}
  Solutions of equation \eqref{eq:Bondi_eq_iso} exist provided that
  $\gmin \leq \fmin - \Lambda$;
  $\gmin=1/2$ occurs for $\Mmin=1$, while $\fmin=2-\ln 2$ is reached at 
  $\xmin=1/2$. Therefore, in the isothermal case, 
  \begin{equation}
        \lambdaup \leq \lcr \equiv \e^{\hspace{0.5mm}\fmin-\,\gmin} = \frac{\e^{3/2}}{4}\hspace{0.2mm},
  \label{eq:lcr_classic}
  \end{equation}
  in agreement with the limit of equation \eqref{eq:lambda_cr} for $\gamma \to 1$.
  
  \subsection{Bondi accretion with electron scattering in galaxy models}\label{subsec:Bondi_gal}
  
  For future use, we now resume the framework used in the previous works
  (KCP16; CP17; CP18) to discuss the Bondi accretion onto MBHs at the centre
  of galaxies, also in presence of radiation pressure due to electron scattering
  (see e.g. Taam, Fu \& Fryxell 1991; Fukue 2001; Lusso \& Ciotti 2011;
  Raychaudhuri, Ghosh \& Joarder 2018; Samadi, Zanganeh \& Abbassi 2019;
  Ram\'{i}rez-Velasquez et al. 2019), 
  and including the additional gravitational field of the galaxy.
  The radiation feedback, in the optically thin regime, can be implemented as a 
  reduction of the gravitational force of the MBH by the factor
  \begin{equation}
        \chiup \equiv 1-\frac{L}{\Ledd}\hspace{0.2mm},
        \qquad\,\,\,
        \Ledd=\frac{4\upi c\hspace{0.3mm}G\MBH\hspace{0.1mm}\mH}{\sigmaThom}\hspace{0.2mm},
  \end{equation}
  where $L$ is the accretion luminosity, $\Ledd$ is Eddington's 
  luminosity, $c$ is the speed of light in vacuum, and 
  $\sigmaThom=6.65 \times 10^{-25}\;{\rm cm}^2$ 
  is the Thomson cross section.
  The (relative) gravitational potential of the galaxy, in general, can be written as
  \begin{equation}
        \Psig=\frac{G\Mg}{\rg}\hspace{0.4mm}\psi\!\left(\frac{r}{\rg}\right)\hspace{-0.1mm},
  \end{equation}
  where $\rg$ is a characteristic scale length of the galaxy density distribution
  (stars plus dark matter), $\psi$ is the dimensionless galaxy potential,
  and finally $\Mg$ is the total mass of the galaxy. 
  For galaxies of infinite total mass, as the J3 models, $\Mg=\Rg\Ms$ is a mass scale
  (see equations \eqref{eq:rhos} and \eqref{eq:Ms&Psis}).
  By introducing the two parameters
  \begin{equation}
        \MR \equiv \frac{\Mg}{\MBH}\hspace{0.2mm},
        \qquad\quad
        \xi \equiv \frac{\rg}{\rB}\hspace{0.2mm},
  \label{eq:R_xi}
  \end{equation}
  where $\rB$ is again defined as in equation \eqref{eq:rB},
  the total relative potential becomes
  \begin{equation}
        \PsiT=\frac{G\MBH}{\rB}\!\left[\hspace{0.3mm}\frac{\chiup}{x}+\frac{\MR}{\xi}\hspace{0.4mm}\psi\!\left(\frac{x}{\xi}\right)\hspace{-0.2mm}\right]\!.
  \label{eq:PsiT_Bondi}
  \end{equation}
  Of course, when $\MR \to 0$ (or $\xi \to \infty$), the 
  galaxy contribution to the total potential vanishes\footnote{For galaxy models
  of finite total mass, or with a total density profile decreasing at large radii at least as
  $r^{-3}$ (as for NFW or King (1972) profiles) $\psi$ can be taken to be zero at infinity 
  (e.g. Ciotti 2021, Chapter 2).}, 
  and the problem reduces to classical case.
  In the limit of $L=\Ledd$ (i.e. $\chiup=0$), the radiation pressure cancels
  the gravitational field of the MBH, then the problem describes
  accretion in the potential of the galaxy only, in absence of electron
  scattering and an MBH; when $L=0$ (i.e. $\chiup=1$), the radiation
  pressure has no effect on the accretion flow.
  Therefore, for MBH accretion in galaxies and in presence of electron scattering, 
  the Bondi problem reduces to the solution of equations 
  \eqref{eq:Bondi_eq_iso} and \eqref{eq:B_iso}, or
  \eqref{eq:Bondi_eq_poly} and \eqref{eq:B_poly},
  where $f$ is now given by
  \begin{equation}
        f(x)=
        \begin{cases}
        \displaystyle \frac{\chiup}{x}+\frac{\MR}{\xi}\hspace{0.4mm}\psi\!\left(\frac{x}{\xi}\right)\hspace{-0.3mm}+2\ln x\hspace{0.2mm},
        \hspace{2.2cm} \gamma=1\hspace{0.1mm},
        \\[15pt]
        \displaystyle\hspace{0.2mm} x^{\frac{4(\gamma-1)}{\gamma+1}}\!\left[\hspace{0.3mm}\frac{\chiup}{x}+\frac{\MR}{\xi}\hspace{0.4mm}\psi\!\left(\frac{x}{\xi}\right)\hspace{-0.3mm}+\frac{1}{\gamma-1}\hspace{0.2mm}\right]\hspace{-0.2mm},
        \hspace{6.2mm} 1<\gamma\leq\frac{5}{3}\hspace{0.2mm},
        \end{cases}
  \label{eq:f_gen}
  \end{equation}
  while the function $g$ (and in particular the value of $\gmin$) is 
  unchanged by the presence of the galaxy.
  Of course, $\Psig$ affects the values of $\xmin$, $\fmin$, and
  of the critical $\lambdaup$ (which now we call $\lt$).
  Two considerations are in order here.
  Firstly, $\Psig$ can produce more than one minimum for the function 
  $f$ (see the case of Hernquist galaxies in CP17); in this circumstance, the
  general considerations after equations \eqref{eq:Bondi_eq_poly} and 
  \eqref{eq:Bondi_eq_iso} force to conclude that $\lt$ is determined by the
  absolute minimum of $f$. Secondly, for a generic galaxy model one cannot
  expect to be able to determine analytically the value of $\xmin$; quite
  surprisingly, in a few cases it has been shown that this is possible
  (see CP18 and references therein). In the following we add another analytical
  case to this list. 

%%%%%%%%%%%%%%%%%%%%%%%%%%%%%%%%%%%%%%%%%%%%%%%%%%%%%%%%%%%%%%%%%%%%%%%%
  \section{The J3 galaxy models}\label{sec:J3}
%%%%%%%%%%%%%%%%%%%%%%%%%%%%%%%%%%%%%%%%%%%%%%%%%%%%%%%%%%%%%%%%%%%%%%%%

  The J3 models (CMP19) are an extension of the  
  JJ models (CZ18), adopted 
  in CP18 to study the isothermal Bondi accretion in two-component
  galaxies with a central MBH. The J3 models are an analytically tractable
  family of spherical models with a central MBH, with a Jaffe (1983) stellar
  density profile, and with a {\it total} density distribution
  such that the DM halo (obtained as the difference between the total and 
  stellar density profiles) is described very well by the NFW 
  profile; in the case of JJ models, instead, the DM profile at large radii declines
  as $r^{-4}$ instead of $r^{-3}$.   
  
  The stellar and total (stars plus DM) density profile of J3 galaxies are 
  then given by
  \begin{equation}
        \rhos(r)=\frac{\rhon}{s^2(1+s)^2}\hspace{0.2mm},
        \qquad\quad
        \rhog(r)=\frac{\Rg\rhon}{s^2(\xig\hspace{-0.15mm}+s)}\hspace{0.2mm},
  \label{eq:rhos}
  \end{equation}
  with
  \begin{equation}
        \rhon=\frac{\Ms}{4\upi\rs^3}\hspace{0.2mm},
        \qquad\,\,\,\,
        s=\frac{r}{\rs}\hspace{0.2mm},
        \qquad\,\,\,\,
        \xig=\frac{\rg}{\rs}\hspace{0.2mm},
  \end{equation}   
  where $\rs$ is the stellar scale length, $\Ms$ is the total stellar 
  mass, $\rg$ is the galaxy scale length, and $\Rg$ measures the 
  total-to-stellar density; for example, we recall that $\Rg/\xig$ gives 
  the ratio $\rhog/\rhos$ for $r \to 0$. The effective radius $\Reff$ of the Jaffe 
  profile is $\Reff \simeq 0.75\,\rs$. The stellar and total mass 
  profiles read
  \begin{equation}
        \Ms(r)=\Ms\,\frac{s}{1+s}\hspace{0.2mm},
        \qquad\,\,\,
        \Mg(r)=\Ms\Rg\ln\frac{\xig\hspace{-0.1mm}+s}{\xig}\hspace{0.2mm},
  \label{eq:Ms&Psis}
  \end{equation}
  so that $\Mg(r)$ diverges logarithmically for $r \to \infty$.
  
  %%%%%%%%%%%%%%%%%%%%%%%%%%%%%%%%%%%%%%%%%%%%%%%%%%%%%%%%%%%%%%%%%%%%%%%%%%%%%%
  \begin{figure*}
        \centering
        \includegraphics[width=0.48\linewidth]{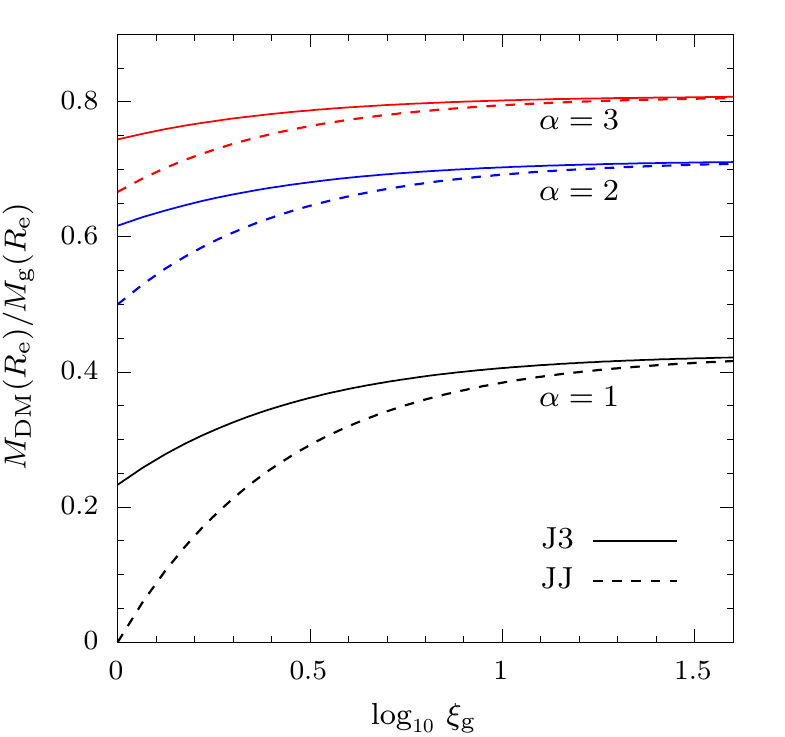}
        \quad\,\,\,
        \includegraphics[width=0.48\linewidth]{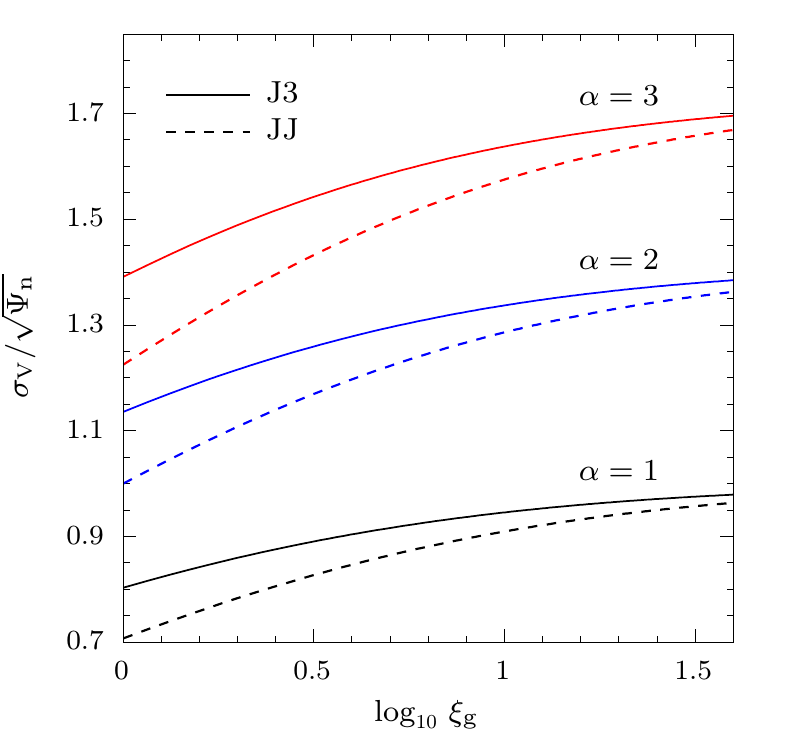}
        \caption{Left: dark-to-total mass ratio of the J3 models (solid lines) within a 
                 sphere of radius $r=\Reff \simeq 0.75\,\rs$ as a function of 
                 $\xig=\rg/\rs$ for the minimum halo case $\alpha=1$ (black), 
                 $\alpha=2$ (blue), and $\alpha=3$ (red). 
                 For comparison, the analogous curves in the case of JJ 
                 models (dashed lines) are shown.
                 Right: the galactic virial velocity dispersion $\sigV$ (solid lines) as 
                 a function of $\xig$, for  
                 $\alpha=1$ (black), $\alpha=2$ (blue), and $\alpha=3$ (red). For 
                 comparison, the analogous curves in the case of JJ models (dashed 
                 lines) are shown.}
  \label{fig:MDM_sigV}
  \end{figure*}
  %%%%%%%%%%%%%%%%%%%%%%%%%%%%%%%%%%%%%%%%%%%%%%%%%%%%%%%%%%%%%%%%%%%%%%%%%%%%%%
  
  The DM halo density profile is therefore
  \begin{equation}
        \rhoDM(r)\equiv\rhog(r)-\rhos(r)
                 =\frac{\rhon}{s^2}\hspace{-0.3mm}
                  \left[\hspace{0.3mm}
                        \frac{\Rg}{\xig\hspace{-0.1mm}+s}-
                        \frac{1}{(1+s)^2}
                        \hspace{0.3mm}
                  \right]\!,
  \label{eq:rhoDM}
  \end{equation}
  and, as shown in CMP19, the condition for a nowhere negative $\rhoDM$ is 
  \begin{equation}
        \Rg \geq \Rm \equiv
            \begin{cases}
              \displaystyle{\frac{1}{4\hspace{0.1mm}(1-\xig)}}\hspace{0.2mm},
              \hspace{5mm} 0 < \xig \leq \frac{1}{2}\hspace{0.2mm},
              \\[10pt]
              \displaystyle{\hspace{0.25mm}\xig}\hspace{0.4mm},
              \hspace{1.59cm} \xig \geq \frac{1}{2}.
            \end{cases}
  \label{eq:pos_cond}
  \end{equation}
  A model with $\Rg=\Rm$ is called a {\it minimum halo} model.
  For assigned $\xig$, it is convenient to introduce the parameter 
  $\alpha$, defined as
  \begin{equation}
        \Rg=\alpha\hspace{0.2mm}\Rm\hspace{0.4mm},
        \qquad\quad
        \alpha \geq 1,
  \label{eq:alpha}
  \end{equation} 
  and, as we shall restrict to the natural situation $\xig \geq 1$, in
  the following $\Rg=\alpha\hspace{0.2mm}\xig$, with $\alpha=1$ corresponding
  to the minimum halo model. Therefore, from equations \eqref{eq:Ms&Psis}, 
  the relative amount of dark-to-total mass as a function of radius is
  \begin{equation}
        \frac{\MDM(r)}{\Mg(r)}=
        1-\frac{s}{\alpha\hspace{0.2mm}\xig(1+s)\hspace{-0.35mm}\ln\hspace{0.2mm}(1+s/\xig)}\hspace{0.2mm},
  \label{eq:MDM}
  \end{equation}
  where $\MDM(r)=\Mg(r)-\Ms(r)$. In Fig. \ref{fig:MDM_sigV} (left panel, solid lines) we 
  plot equation \eqref{eq:MDM} as a function of $\xig \geq 1$ for $r=\Reff$  
  and for three values of $\alpha$\hspace{0.1mm}: the
  minimum halo model ($\alpha=1$) and two cases with $\alpha>1$. Fractions
  of DM with values for the minimum 
  halo case in agreement with those required by the dynamical modelling of 
  early-type galaxies (see e.g. Cappellari et al. 2015) can be easily obtained. 
  These fractions are unsurprisingly sligthly larger than those obtained in the 
  case of JJ models for the same values of $\xig$ (see 
  dashed lines in Fig. \ref{fig:MDM_sigV}, left panel).
  
  Notice that by construction, $\rhoDM \propto r^{-3}$ at large radii, while,
  as in JJ models, at small radii
  $\rhoDM \propto r^{-2}$ (i.e. the DM and stellar densities are locally 
  proportional), with the exception of the minimum halo models, in which 
  $\rhoDM \propto r^{-1}$. We now compare the DM profile of J3 models with 
  the untruncated NFW profile (Navarro et al. 1997), that in our notation can be written as
  \begin{equation}
        \rhoNFW(r)=\frac{\rhon\RNFW}{q(c)s\hspace{0.2mm}(\xiNFW+s)^2}\hspace{0.2mm},
        \quad
        q(c) \equiv \ln\hspace{0.2mm}(1+c)-\frac{c}{1+c}\hspace{0.2mm},
  \label{eq:rhoNFW}
  \end{equation}
  where $\xiNFW \equiv \rNFW/\rs$ is the NFW scale length in units of $\rs$
  and, for a chosen reference radius $\rt$, we define $\RNFW \equiv \MNFW(\rt)/\Ms$ 
  and $c \equiv \rt/\rNFW$. The densities $\rhoDM$ and $\rhoNFW$
  can be made asymptotically identical both at small {\it and} large radii
  by fixing
  \begin{equation}
        \RNFW=q(c)\hspace{0.4mm}\xig\hspace{0.4mm},
        \qquad\quad
        \xiNFW=\frac{\xig}{\sqrt{\hspace{0.3mm}2\hspace{0.3mm}\xig-1}}\hspace{0.2mm}.
  \label{eq:NFW_conds}
  \end{equation}
  Hence, once a specific minimum halo galaxy model is 
  considered, equations \eqref{eq:rhoNFW} and 
  \eqref{eq:NFW_conds} allow to determine the NFW profile that best reproduces 
  the DM halo density profile. 
  Cosmological simulations suggest for galaxies $c\simeq 10$ (see e.g.
  Bullock \& Boylan-Kolchin 2017), and $\RNFW \simeq $ a few tens. 
  Moreover, the value of $\xig$ cannot be too large, otherwise the DM fraction 
  inside $\Reff$ would exceed the values derived from observations (see e.g. Napolitano et al. 2010;
  see also Fig. 1 in CMP19). For these reasons we conclude
  that {\it the} NFW {\it shape and the cosmological expectations 
  are reproduced if we consider minimum halo models with} $\xig \simeq 10 \div 20$.
  In the following, we choose as `reference model' a minimum halo model with $\Rg=\xig=13$,
  $c=10$, $\RNFW \simeq 20$, and $\rNFW=2.6\,\rs$.
  
  \subsection{Central and Virial properties of J3 models}\label{subsec:Centr_Vir}
  
  Now we recall a few dynamical properties of the J3 models needed in the following
  discussion (see CMP19 for more details).
  A MBH of mass $\MBH=\mu\Ms$ is added at the centre of the galaxy, and the total 
  (relative) potential is
  \begin{equation}
        \PsiT(r)=\frac{\Psin\mu}{s}+\Psig(r)\hspace{0.4mm},
        \quad\,\,\,
        \Psin=\frac{G\Ms}{\rs}\hspace{0.2mm},
        \quad\,\,\,
        \mu=\frac{\MBH}{\Ms}\hspace{0.2mm},
  \end{equation}
  where
  \begin{equation}
        \Psig(r)=\frac{\Psin\Rg}{\xig}
                 \left(
                 \hspace{0.25mm}
                      \ln\frac{\xig\hspace{-0.1mm}+s}{s}+
                      \frac{\xig}{s}\ln\frac{\xig\hspace{-0.1mm}+s}{\xig}
                 \hspace{0.25mm}
                 \right)\hspace{-0.2mm};
  \label{eq:Psig}
  \end{equation}
  in particular, $\Psig \propto (\ln s)/s$ at large radii, and $\Psig \propto -\ln s$
  near the centre. 
  The stellar orbital structure is limited to the isotropic case. 
  The radial component of the velocity dispersion is given by
  \begin{equation}
        \sigr^2(r)=\sigBH^2(r)+\sigg^2(r)\hspace{0.4mm},
  \end{equation}
  where $\sigBH$ and $\sigg$ indicate, respectively, the contribution of 
  the MBH and of the galaxy potential. As shown in CMP19, the Jeans equations
  for J3 models can be solved analytically, here we just recall that
  in the isotropic case
  \begin{equation}
        \sigr^2(r) \sim \Psin\hspace{-0.2mm}\times
        \begin{cases}
        \displaystyle \frac{\mu}{3\hspace{0.1mm}s}
                        +\frac{\Rg}{2\hspace{0.2mm}\xig}-\frac{\mu}{3}\hspace{0.2mm},
        \hspace{8.5mm} r \to 0\hspace{0.2mm},
        \\[12pt]
        \displaystyle \Rg\hspace{0.2mm}\frac{\ln s}{5\hspace{0.1mm}s}\hspace{0.2mm},
        \hspace{1.78cm} r \to \infty\hspace{0.2mm},
        \end{cases}
  \label{eq:sigr_center_inf}
  \end{equation}
  where, for mathematical consistency, we retained also the constant term
  $-\hspace{0.2mm}\mu/3$ in the asymptotic expansion of $\sigr$ near the centre, although
  this contribution is fully negligible in realistic galaxy models.
  Notice that, when $\xig \geq 1$, from equation \eqref{eq:alpha} it follows that
  the constant term due to the galaxy is independent of $\xig$, with
  $\sigg^2(0)=\Psin\alpha/2$. This latter expression provides the 
  interesting possibility of adopting $\sigg(0)$ as a proxy for the observed velocity 
  dispersion of the galaxy in the central regions, outside the sphere of influence of the 
  central MBH.
  
  In order to derive an estimate of the sphere of influence of the MBH,
  it is interesting to consider the projected velocity dispersion 
  $\sigp(R\hspace{0.3mm})=\sqrt{\hspace{0.3mm}\sigpBH^2(R\hspace{0.3mm})+\sigpg^2(R\hspace{0.3mm})\hspace{0.2mm}}\hspace{0.2mm}$, 
  where $R$ is the radius in the 
  projection plane. At large radii $\sigp$ is dominated
  by the galaxy contribution: from equation \eqref{eq:sigr_center_inf} one
  has, at the leading order,
  \begin{equation}
        \sigp^2(R\hspace{0.3mm}) \sim \frac{\Psin\hspace{0.15mm}8\hspace{0.2mm}\Rg\hspace{-0.2mm}\ln\eta}{15\upi\eta}\hspace{0.2mm},
        \qquad
        \eta \equiv \frac{R}{\rs}.
  \end{equation}
  At small radii, instead (CMP19, equations (57) and (58)), 
  \begin{equation}
        \sigpg(0)=\sigg(0)=\frac{\Psin\Rg}{2\hspace{0.2mm}\xig}\hspace{0.2mm},
        \qquad\,\,\,\,\,
        \sigpBH^2(R\hspace{0.3mm}) \sim \frac{\Psin\hspace{0.15mm}2\hspace{0.2mm}\mu}{3\upi\eta}\hspace{0.1mm}.
  \label{eq:sigpg=sigg}
  \end{equation}
  Equation \eqref{eq:sigpg=sigg} allows to estimate
  the radius $\Rinf$ of the sphere of influence, defined
  as the distance from the centre in the projection plane where $\sigp$ in presence of the MBH 
  exceedes by a factor $(1+\epsilon)$ the galaxy projected velocity dispersion $\sigpg$
  in absence of the MBH: 
  \begin{equation}
        \sqrt{\hspace{0.4mm}\sigpBH^2(\Rinf)+\sigpg^2(\Rinf)\hspace{0.4mm}}\hspace{0.5mm} 
        \equiv 
        \hspace{0.2mm}(1+\epsilon)\hspace{0.2mm}\sigpg(\Rinf)\hspace{0.3mm}.
  \label{eq:Rinf_def}
  \end{equation}
  In practice, for a galaxy model with finite $\sigpg(0)$, the formula above reduces
  to equation (36) in CP18, and for $\xig \geq 1$ equation \eqref{eq:sigpg=sigg}
  yields
  \begin{equation}
        \frac{\Rinf}{\rs}
        \simeq\frac{4\mu}{3\upi\alpha\hspace{0.2mm}\epsilon\hspace{0.2mm}(2+\epsilon)}\hspace{0.2mm}.
  \label{eq:Rinf}
  \end{equation}
  Notice that equation \eqref{eq:Rinf} is
  coincident with the same estimate in JJ models (CP18, equation 37), being the 
  two models identical in the central regions.
  
  A fundamental ingredient in Bondi accretion is the gas temperature at infinity $\Tinf$.
  As in CP18, in the next Section we shall use 
  $\TV\!=\!\mugas\hspace{0.1mm}\mH\hspace{0.2mm}\sigV^2/(3\hspace{0.2mm}\kB)$ 
  (see e.g. Pellegrini 2011) as the natural scale for $\Tinf$,
  where $\sigV$ is the (three dimensional) virial velocity dispersion of stars obtained
  from the Virial Theorem:
  \begin{equation}
         \Ms\sigV^2 \equiv 2\Ks = -\hspace{0.7mm}\Wsg-\WsBH\hspace{0.2mm}.
  \label{eq:VT}
  \end{equation}
  In the equation above, $\Ks$ is the total kinetic energy of the stars,
  \begin{equation}
        \Wsg=-\,4\upi G\!\int_0^{\infty}\!\Mg(r)\rhos(r)\hspace{0.3mm}rdr
  \end{equation}
  is the interaction energy of the stars with the gravitational field of
  the galaxy (stars plus DM), and the MBH contribution $\WsBH$
  diverges near the origin for a Jaffe density distribution.
  Since we shall use $\sigV$ as a proxy for the gas temperature at large distance
  from the centre, we neglect $\WsBH$ in equation \eqref{eq:VT}, so that
  \begin{equation}
        \sigV^2=-\hspace{0.4mm}\frac{\Wsg}{\Ms}=\Psin\Rg\tWsg\hspace{0.2mm},
        \qquad
        \tWsg=\Hf(\xig,0)-\frac{\ln\xig}{\xig-1}\hspace{0.2mm}, 
  \label{eq:sigV}
  \end{equation}
  where the function $\Hf(\xig,s)$ is given in Appendix C of CMP19.
  Fig. \ref{fig:MDM_sigV} (right panel) shows the trend of $\sigV$ as a function of $\xig \geq 1$,
  for three J3 (solid) and JJ (dashed) models. As expected, $\sigV$
  increases with $\xig$, and
  $\sigV \simeq \sqrt{\alpha\hspace{0.1mm}\Psin}$ when $\rg \gg \rs$.
  For comparison, we show in Fig. \ref{fig:MDM_sigV} (right panel) $\sigV$  
  for the JJ models of same parameters.
  
  %%%%%%%%%%%%%%%%%%%%%%%%%%%%%%%%%%%%%%%%%%%%%%%%%%%%%%%%%%%%%%%%%%%%%%%%%%%%%%%%%%%%%%%%%%%%%
\begin{table*}
\centering
    \caption{Galaxy Structure and Accretion Flow parameters}
    \begin{tabular}{@{}l@{\hskip 15.2em}r@{\hskip 3em}l@{\hskip 15.6em}r@{}}
        \toprule[1.25pt]\midrule[0.3pt]
        \multicolumn{2}{c@{\hskip 3.25em}}{Galaxy Structure} & \multicolumn{2}{c}{Accretion Flow}\\
        \cmidrule{1-2}\cmidrule{3-4}
        \multicolumn{1}{@{}l}{Symbol} & \multicolumn{1}{r@{\hskip 3em}}{Quantity} & \multicolumn{1}{@{}l}{Symbol} &
        \multicolumn{1}{r@{}}{Quantity} \\
        \midrule
        \vspace{1pt}
        $\Ms$ & Total stellar mass & $\Tinf$ & Gas temperature at infinity \\
        \vspace{1pt}
        $\rs$ & Stellar density scale length & $\rhoinf$ & Gas density at infinity \\ 
        \vspace{1pt}
        $\Mg$ & Total$^{\hspace{0.25mm}\rm a}$\tnote{a} galaxy mass & $\cinf$ & Speed of sound at infinity \\
        \vspace{1pt}
        $\rg$ & Total density scale length & $\gamma$ & Polytropic index ($1 \leq \gamma \leq 5/3$) \\
        \vspace{1pt}
        $\MBH$ & Central MBH mass & $\gammaAD$ & Adiabatic index ($\hspace{0.15mm}=\cp/\cV$) \\
        \vspace{1pt}
        $\mu$ & $\MBH/\Ms$ & $\MR$ & $\Mg/\MBH$ ($\hspace{0.15mm}=\hspace{-0.2mm}\Rg/\mu$) \\
        \vspace{1pt}
        $\Rg$ & $\Mg/\Ms$ ($\hspace{0.15mm}=\alpha\hspace{0.1mm}\Rm\hspace{0.2mm}$) & $\beta$ & $\Tinf/\TV$ \\
        \vspace{1pt}
        $\Rm$ & Minimum value of $\Rg$ & $\rB$ & Bondi radius \\
        \vspace{1pt}
        $\xig$ & $\rg/\rs$ & $\rmin$ & Sonic radius \\
        \vspace{1pt}
        $s$ & $r/\rs$ & $x$ & $r/\rB$ \\   
        \vspace{1pt}
        $\sigV$ & Stellar virial velocity dispersion & $\xi$ & $\rg/\rB$ \\
        \vspace{1pt}
        $\TV$ & Stellar virial temperature & $\lt$ & Critical accretion parameter \\
        \vspace{1pt}
        $\Wsg$ & Virial energy of stars & $\M$ & Mach number \\
     \bottomrule
    \end{tabular}
    \vspace{1.5mm}
\begin{tablenotes}[para,flushleft]
      \item[a] For example, from our definition $\Mg=\Rg\Ms$, and equation \eqref{eq:rhos},
               $\Mg$ is the total mass (stellar plus DM) inside a sphere of radius 
               $(\e-1)\hspace{0.3mm}\rg$.
    \end{tablenotes}
\label{tab:parameters}
\end{table*}
%%%%%%%%%%%%%%%%%%%%%%%%%%%%%%%%%%%%%%%%%%%%%%%%%%%%%%%%%%%%%%%%%%%%%%%%%%%%%%%%%%%%%%%%%%%%
  
%%%%%%%%%%%%%%%%%%%%%%%%%%%%%%%%%%%%%%%%%%%%%%%%%%%%%%%%%%%%%%%%%%%%%%%%
  \subsection{Linking Stellar Dynamics to Fluid Dynamics}\label{sec:Combo}
%%%%%%%%%%%%%%%%%%%%%%%%%%%%%%%%%%%%%%%%%%%%%%%%%%%%%%%%%%%%%%%%%%%%%%%%

  We now link the stellar dynamical properties of the 
  galaxy models with the defining parameters of Bondi accretion introduced
  in Section \ref{subsec:Bondi_gal}.
  In fact, the function $f$ in equation
  \eqref{eq:f_gen} is written in terms of quantities referring to the central 
  MBH and to the gas temperature at infinity, while the stellar 
  dynamical properties of the J3 models are written in terms of the
  observational properties of the galaxy stellar component.
  The two groups of parameters are summarised in Table 2.
  
  The first accretion parameter we consider is $\MR$ in equation \eqref{eq:R_xi}. 
  It is linked to the galaxy structure by the following expression
  \begin{equation}
        \MR \equiv \frac{\Mg}{\MBH}
        =\frac{\Rg}{\mu}
        =\frac{\alpha\hspace{0.3mm}\xig}{\mu}\hspace{0.2mm},
  \label{eq:R}
  \end{equation}
  where the last identity derives from equation \eqref{eq:alpha} with
  $\xig \geq 1$; notice that $\MR \approx 10^4$ for $\xig$ of the order of tens 
  and $\alpha$ of order unity, and $\mu=0.002$
  (see Kormendy \& Ho 2013 for this choice of $\mu$).
  
\begin{figure*}
        %\centering
        \includegraphics[width=0.48\linewidth]{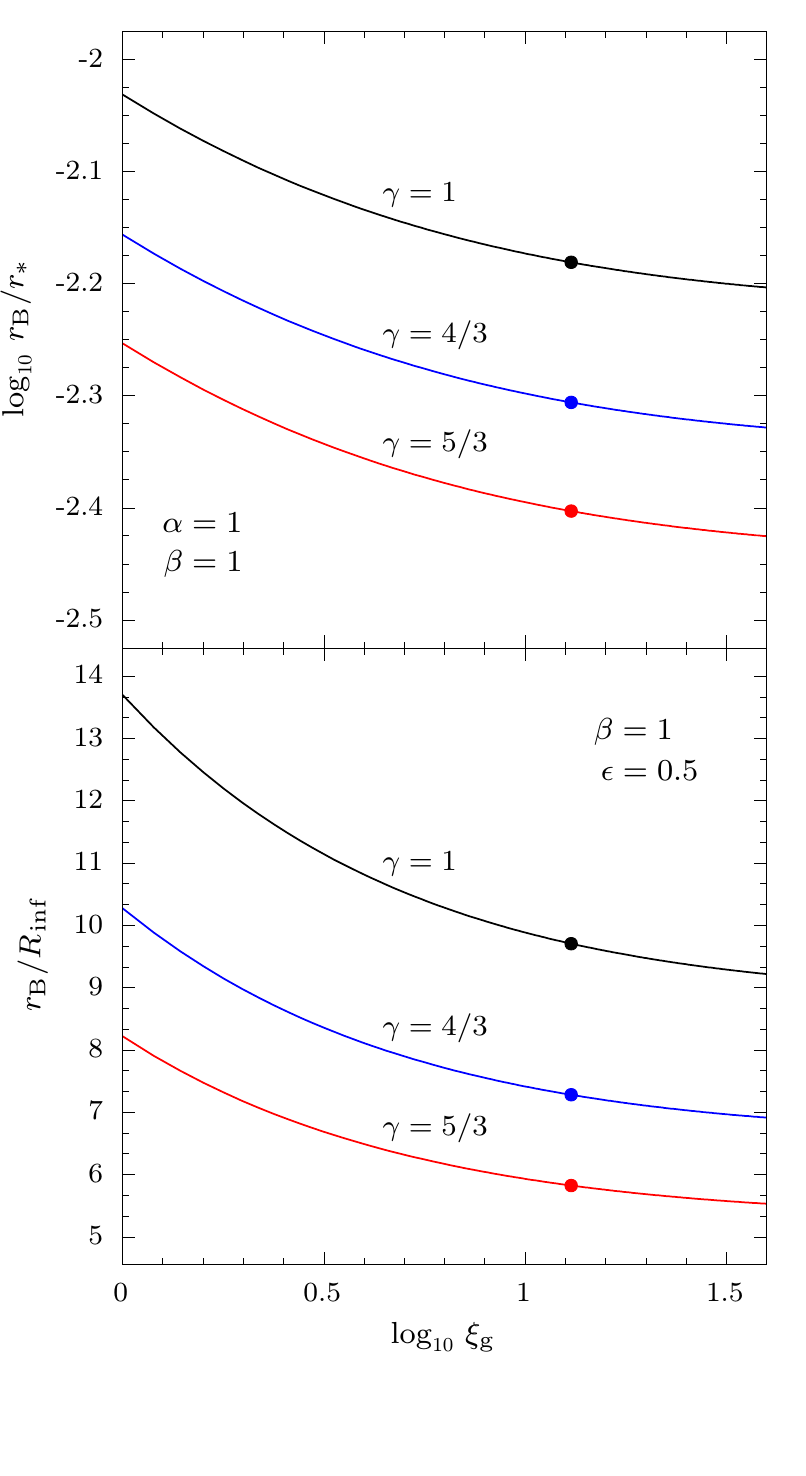}
        \quad\,\,\,
        \includegraphics[width=0.48\linewidth]{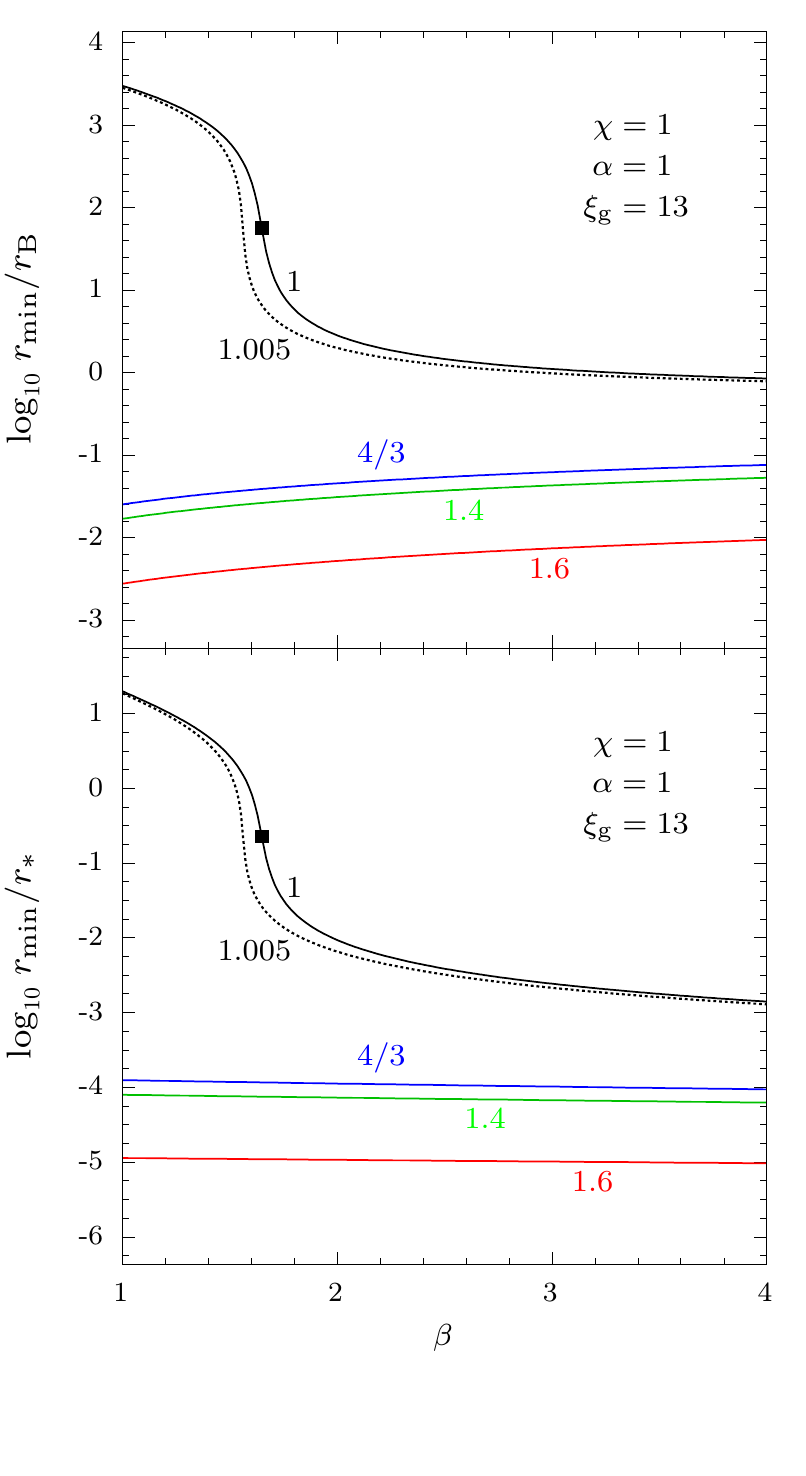}\\
        \vspace{-12mm}
        \caption{Left: Bondi radius $\rB$ in units of $\rs$ (top) and $\Rinf$ (bottom), as a function of the 
                 galaxy-to-stellar scale length ratio $\xig=\rg/\rs$, for three 
                 J3 galaxy models with $\beta \equiv \TV/\Tinf=1$, $\mu=0.002$, and
                 $\gamma=1$, $4/3$, $5/3$;
                 the solid dots at $\rB/\rs \simeq 0.0066$, $0.0049$, $0.0039$ 
                 correspond to the minimum halo case with $\xig=13$;
                 solid dots at $\rB/\Rinf \simeq 9.71$, $7.28$, $5.82$ correspond 
                 to the case $\xig=13$ and $\epsilon=0.5$.
                 Right: position of $\xmin \equiv \rmin/\rB$ (top) and 
                 $\rmin/\rs$ (bottom) as a function of  
                 $\beta=\Tinf/\TV$, in the case of minimum halo models with
                 $\xig=13$, $\mu=0.002$, and $\chiup=1$, for different values of the 
                 polytropic index given close to the curves.
                 The black square points at $\rmin/\rB \simeq 57.34$ and 
                 $\rmin/\rs \simeq 0.23$ correspond to the critical case
                 $\beta=\bc \simeq 1.65$.}
  \label{fig:rmin}
  \end{figure*}

  The determination of the accretion parameter $\xi$ is more articulated. This quantity depends on
  the Bondi radius $\rB$; we stress again that in the present discussion,
  even in presence of the galaxy gravitational potential, $\rB$
  is still defined in the classical sense, i.e., just considering the mass of
  the MBH, as in equation \eqref{eq:rB}.
  Of course, $\rB$ depends on the gas temperature at infinity.
  In principle, arbitrary values of $\Tinf$
  could be adopted, but in real systems the natural scale for the global temperature
  is represented by the virial temperature 
  $\TV$ defined via the virial velocity dispersion in equation \eqref{eq:VT}.
  Accordingly, we set 
  \begin{equation}
        \Tinf=\beta\hspace{0.3mm}\TV,
        \qquad
        \cinf^2=\gamma\hspace{0.4mm}\frac{\pinf}{\rhoinf}
               =\frac{\gamma\beta\hspace{0.1mm}\sigV^2}{3}.
  \end{equation}
  From equations \eqref{eq:rB} and \eqref{eq:sigV} we then obtain
  \begin{equation}
        \frac{\rB}{\rs}=\frac{3\mu}{\alpha\beta\gamma\hspace{0.1mm}\Fg(\xig)}\hspace{0.3mm},
        \qquad\quad
        \Fg \hspace{-0.2mm}\equiv \xig\hspace{0.2mm}\tWsg(\xig)\hspace{0.3mm},
  \label{eq:rB/rs}
  \end{equation} 
  where the function $\Fg$ monothonically increases with $\xig$ from
  $\Fg(1)=\upi^2\hspace{-0.45mm}/\hspace{0.15mm}6-1$ to $\Fg(\infty)=1$.
  For example, at fixed $\alpha$, $\beta$, and $\gamma$, one has 
  \begin{equation}
        \frac{3\mu}{\alpha\beta\gamma}<\frac{\rB}{\rs}\leq\frac{18\mu}{(\upi^2\hspace{-0.4mm}-6)\hspace{0.1mm}\alpha\beta\gamma}.
  \label{eq:rBrs_extremes}
  \end{equation} 
  In Fig. \ref{fig:rmin} (top left) we show the trend of $\rB/\rs$ as
  a function of $\xig$ in the minimum halo case ($\alpha=1$) with $\beta=1$ and $\mu=0.002$, for 
  three values of $\gamma$; in general, $\rB$ is of the order of a few 
  $\times\hspace{0.5mm}10^{-3}\hspace{0.3mm}\rs$. Note that, for fixed $\xig$,
  the isothermal profile (black line) is above that in the corresponding adiabatic case
  (red line); in general, for fixed $\alpha$, $\beta$ and $\xig$, 
  $\rB/\rs$ always lies between the isothermal
  and the monoatomic adiabatic case, as shown by equation \eqref{eq:rBrs_extremes}.
  Finally, by combining equations \eqref{eq:R_xi} and 
  \eqref{eq:rB/rs},
  \begin{equation}
        \xi \equiv\frac{\rg}{\rB}
        =\frac{\alpha\beta\gamma\hspace{0.2mm}\xig\Fg(\xig)}{3\mu}
        =\frac{\MR\beta\hspace{0.2mm}\gamma\Fg(\xig)}{3}\hspace{0.2mm},
  \label{eq:csib}
  \end{equation}
  and so $\MR$ and $\xi$ increase with $\xig$. Curiously, 
  from the general definitions in equation \eqref{eq:R_xi}, and making use
  of equation \eqref{eq:sigpg=sigg},
  \begin{equation}
        \frac{\MR}{\xi}=\frac{2\hspace{0.3mm}\sigpg^2(0)}{\cinf^2},   
  \label{eq:R/xi}       
  \end{equation}
  which links directly the parameters of Bondi accretion to the observable
  $\sigpg(0)=\sigg(0)$.
  
  For observational purposes, it is also useful to express the position of $\rB$
  in terms of the radius $\Rinf$ as given in equation \eqref{eq:Rinf};
  since the parameter $\alpha=\Rg/\Rm$ cancels out, we have 
  \begin{equation}
        \frac{\rB}{\Rinf}=\frac{9\upi\hspace{0.1mm}\epsilon\hspace{0.2mm}(2+\epsilon)}{4\hspace{0.1mm}\beta\gamma\Fg(\xig)}\hspace{0.2mm},
  \end{equation}
  independently of the minimum halo assumption.  
  In Fig. \ref{fig:rmin} (bottom left panel) we show the trend of $\rB/\Rinf$ when $\epsilon=0.5$
  and $\beta=1$, for the same three values of $\gamma$ as in the upper left panel:
  $\rB\approx$ a few times $\Rinf$.
  
  Now we move the discussion to the sonic radius $\rmin$, one of the most 
  important properties of the accretion solution. The effects of the galaxy do indeed
  manifest themselves in the position of $\rmin$.
  When measured in terms of the scale length $\rs$, it can
  be written, by making use of equation \eqref{eq:rB/rs}, as 
  \begin{equation}
        \frac{\rmin}{\rs}=\hspace{0.6mm}\xmin(\chiup,\MR,\xi)\hspace{0.4mm}\frac{\rB}{\rs}\hspace{0.2mm},
  \label{eq:rminrs}
  \end{equation}
  where $\xmin \equiv \rmin/\rB$ gives the (absolute) minimum of $f$. 
  
  Finally, we must recast the galaxy potential in equation \eqref{eq:Psig}
  by using the normalization scales in equation \eqref{eq:PsiT_Bondi}\hspace{0.1mm}:
  as $s/\xig=x/\xi$, it is immediate that in our problem 
  \begin{equation}
        \psi\!\left(\frac{x}{\xi}\right)\!=
        \ln\!\left(1+\frac{\xi}{x}\right)\hspace{-0.2mm}
        +\hspace{0.25mm}\frac{\xi}{x}\ln\!\left(1+\frac{x}{\xi}\right)\hspace{-0.3mm}.
  \label{eq:psi(x/csi)}
  \end{equation}  

%%%%%%%%%%%%%%%%%%%%%%%%%%%%%%%%%%%%%%%%%%%%%%%%%%%%%%%%%%%%%%%%%%%%%%%%%%%%%%
  %%%%%%%%%%%%%%%%%%%%%%%%%%%%%%%%%%%%%%%%%%%%%%%%%%%%%%%%%%%%%%%%%%%%%%%%%%%%%%

%%%%%%%%%%%%%%%%%%%%%%%%%%%%%%%%%%%%%%%%%%%%%%%%%%%%%%%%%%%%%%%%%%%%%%%%%
  \section{Bondi accretion in J3 models}\label{sec:BondiJ3}
%%%%%%%%%%%%%%%%%%%%%%%%%%%%%%%%%%%%%%%%%%%%%%%%%%%%%%%%%%%%%%%%%%%%%%%%%
  
  We can now discuss the full problem, investigating how the standard 
  Bondi accretion is modified by the additional potential of J3 galaxies, 
  and by electron scattering. We show that in the isothermal case ($\gamma=1$)
  the solution is fully analytical, as for the monoatomic adiabatic case
  ($\gamma=5/3$); for $1<\gamma<5/3$, instead,
  it is not possible to obtain analytical expressions, and so 
  a numerical investigation is presented.
  
%%%%%%%%%%%%%%%%%%%%%%%%%%%%%%%%%%%%%%%%%%%%%%%%%%%%%%%%%%%%%%%%%%%%%%%%%
  \subsection{The $\gamma=1$ case}\label{sec:J3poly}
%%%%%%%%%%%%%%%%%%%%%%%%%%%%%%%%%%%%%%%%%%%%%%%%%%%%%%%%%%%%%%%%%%%%%%%%%

 %%%%%%%%%%%%%%%%%%%%%%%%%%%%%%%%%%%%%%%%%%%%%%%%%%%%%%%%%%%%%%%%%%%%%%%%%%%%%%
  \begin{figure*}
        %\centering
        \includegraphics[width=0.48\linewidth]{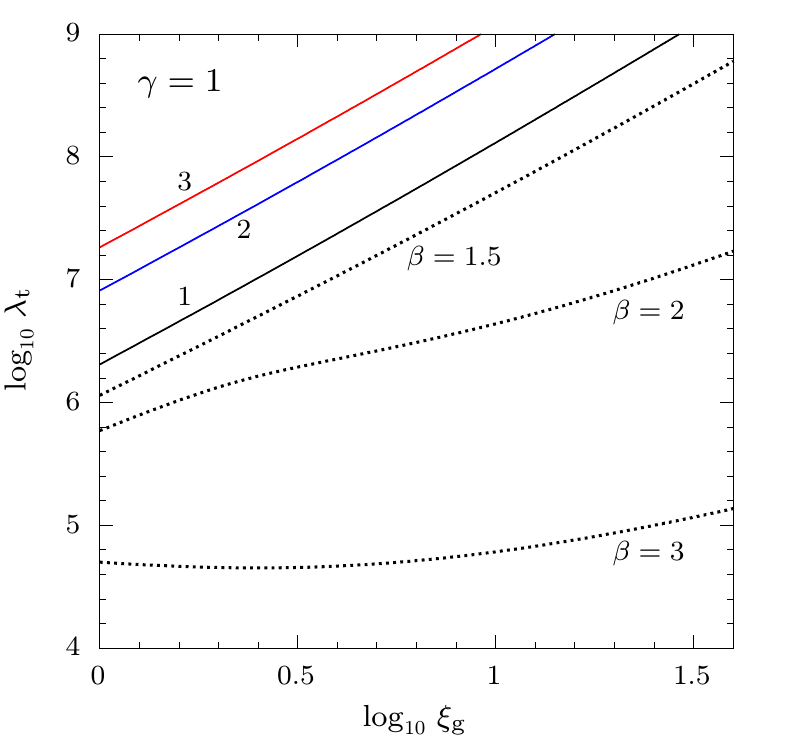}
        \quad\,\,\,
        \includegraphics[width=0.48\linewidth]{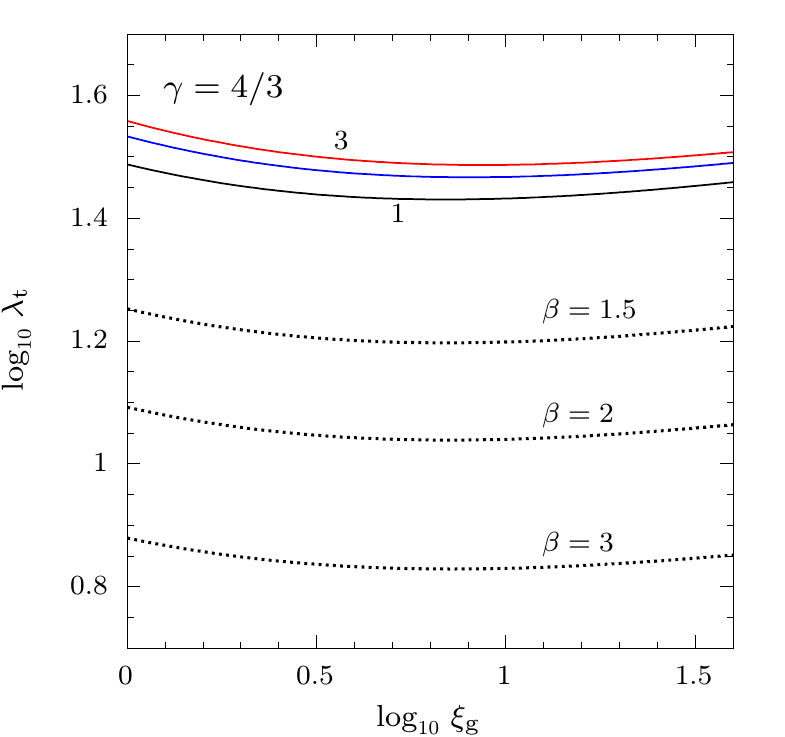}\\
        %\vspace{1mm}
        \caption{Critical accretion parameter $\lt$ as a function of $\xig$, for the 
                 minimum halo case $\alpha=1$ (black), $2$ (blue), and $3$ (red), 
                 and $\chiup=1$ and $\beta=1$. The dotted curves refer to $\alpha=1$ 
                 and three different values of $\beta$. The left panel shows the case
                 $\gamma=1$, the right panel shows the case $\gamma=4/3$.
                 Notice that $\lt$ in the isothermal case is several order of magnitude
                 larger than the $\gamma=4/3$ case.}
  \label{fig:lt}
  \end{figure*}
  %%%%%%%%%%%%%%%%%%%%%%%%%%%%%%%%%%%%%%%%%%%%%%%%%%%%%%%%%%%%%%%%%%%%%%%%%%%%%%
  
  The isothermal case stands out not 
  only because $f(x)$ in equation \eqref{eq:f_gen} is not of
  the general family, for $\gamma=1$, but also because the position of the
  sonic radius $\xmin$ can be obtained explicitely. Indeed, for 
  $0<\chiup\leq 1$, $f(x)\sim\chiup/x$ for $x \to 0$, and $\sim 2\ln x$ for $x \to \infty$.
  Therefore, the continuous function $f$ has at least one critical point
  over the range $0 \leq x < \infty$, obtained by solving  
  \begin{equation}
  2\hspace{0.2mm}\xmin-\MR\ln\frac{\xi+\xmin}{\xi}=\chiup\hspace{0.3mm}.
  \label{eq:df/dx_iso}
  \end{equation}
  As shown in Appendix \ref{app:W}, the positive solution can be 
  obtained for generic values of the model parameters in terms of the
  Lambert-Euler $W$ function\footnote{The $W$ function is not new in the study of isothermal flows. 
  See, e.g., Cranmer (2004), Waters \& Proga (2012), Herbst (2015), CP17, CP18.}, and 
  the {\it only} minimum of $f$ is reached at 
  \begin{equation}
  \xmin \equiv \frac{\rmin}{\rB}
        = -\hspace{0.8mm}\xi-\frac{\MR}{2}\,\Wmu\hspace{-0.4mm}\left(-\hspace{0.5mm}\frac{2\xi}{\MR}\hspace{0.4mm}\e^{-\hspace{0.4mm}\frac{\chiup+2\xi}{\MR}}\right)\hspace{-0.3mm}.
  \label{eq:xminISO}
  \end{equation}
  Once $\xmin$ is known, all the other quantities in 
  the Bondi solution, such as the critical 
  accretion parameter $\lt=\exp\hspace{0.25mm}(\fmin\hspace{-0.1mm}-1/2)$ 
  in equation \eqref{eq:lcr_classic}, the mass accretion
  rate in equation \eqref{eq:continuity}, and the Mach number profile
  $\M$, can be expressed as a function of $\xmin$.
  Therefore, J3 galaxies belong to the family of models for which a 
  fully analytical discussion of the isothermal Bondi accretion problem is
  possible (see Table 1).
  In particular, from CP17 and CP18, the critical accretion solution reads
  \begin{equation}
        \M^2=-
             \begin{cases}
                   \displaystyle{\Wz\hspace{0.2mm}\big(\hspace{-0.3mm}-\hspace{-0.2mm}\lt^2\hspace{0.3mm} \e^{-2f}\big)}\hspace{0.3mm}, \hspace{11mm} x \geq \xmin\hspace{0.3mm},
                   \\[7pt]
                   \displaystyle{\Wmu\hspace{0.2mm}\big(\hspace{-0.3mm}-\hspace{-0.2mm}\lt^2\hspace{0.3mm} \e^{-2f}\big)}\hspace{0.3mm}, \hspace{6.5mm} 0 < x \leq \xmin\hspace{0.3mm},
             \end{cases}
  \label{eq:MachISO}
  \end{equation}
  where
  $f$ is given in equation \eqref{eq:f_gen} with the function $\psi(x/\xi)$
  defined by equation \eqref{eq:psi(x/csi)}.  
  Summarising, $\Wmu$ describes supersonic accretion, 
  while $\Wz$ subsonic accretion\footnote{
  As $x$ decreases from $\infty$ to $\xmin$, the argument of  
  $\Wz$ decreases from $0$ to $-\hspace{0.4mm}1/\e$ (points $A$ and $B$ in Fig. 
  \ref{fig:Lambert}, left panel), 
  and $\M^2$ increases from $0$ to $1$.
  As $x$ further decreases from $\xmin$ to $0$, the argument of $\Wmu$ increases 
  again from $-\hspace{0.4mm}1/\e$ to $0$ (points $B$ and $C$), 
  and $\M^2$ increases from $1$ to $\infty$. The other critical solution, with
  $\M^2$ increasing for increasing $x$, is obtained by switching the functions $\Wz$ 
  and $\Wmu$ in equation \eqref{eq:MachISO}.}.
  Although equation \eqref{eq:MachISO} provides an explicit expression of $\M$,
  it can be useful to have its asymptotic 
  trend at small and large distances from the centre; from equation 
  \eqref{eq:Wz_asymp} and the expansion of $f(x)$, one has
  \begin{equation}
        \M^2 \sim
        \begin{cases}
              \displaystyle{\frac{2\chiup}{x}+\frac{2(2\xi-\MR)}{\xi}\ln\frac{x}{\xmin}}\hspace{0.2mm},
              \hspace{0.8cm} x \to 0\hspace{0.4mm},
              \\[12pt]
              \displaystyle{\lt^2\hspace{0.4mm}x^{-\hspace{0.25mm}2\hspace{0.1mm}\left(2\hspace{0.4mm}+\frac{\MR}{x}\hspace{-0.2mm}\right)}}\hspace{0.2mm},
              \hspace{2.273cm} x \to \infty\hspace{0.1mm}.
        \end{cases}
  \label{eq:M2_asymp}
  \end{equation}
  Of course, the same result can be established also by asympotic expansion of 
  equation \eqref{eq:Bondi_eq_iso}.
  Therefore, in the central region $\M \propto x^{-1/2}$ for $\chiup>0$, while 
  $\M \sim \sqrt{\hspace{0.2mm}2(2-\MR/\xi)\hspace{-0.35mm}\ln\hspace{0.2mm}(x/\xmin)}$ 
  when $\chiup=0$ (provided that $\MR>2\xi$).
  
  As already found for JJ models in the isothermal case, also for J3 models the case $\chiup=0$
  (from equation (\ref{eq:PsiT_Bondi}) corresponding to a galaxy 
  without a central MBH) reveals some interesting properties of the
  gas flow, also relevant for the understanding of the more natural situation $\chiup>0$.
  In fact, near the centre $f(x) \sim (2-\MR/\xi)\ln x$, and a solution is possible only for $\MR \geq 2\xi$, 
  with $\xmin$ given by equation \eqref{eq:xminISO}.
  When $\MR<2\xi$, $\fmin=-\,\infty$ (reached at the origin), and therefore no accretion 
  is possible since $\lt$ would be zero. 
  In the special case $\chiup=0$ and $\MR=2\xi$, $\fmin$
  is again reached at the origin, but $f$ now converges to $\fmin=2(1+\ln\xi)$, with $\lt=\xi^2\e^{3/2}$.
  Given the similarity of JJ and J3 models near the centre, the fact that both models
  share the same properties at small radii is not surprising\footnote{For a further
  discussion of the effect of the central density slope on the existence of 
  isothermal accretion solutions with $\chiup=0$, see CP17 and CP18.}.
  Equation \eqref{eq:M2_asymp} can still be used with $\chiup=0$ for
  $\MR>2\xi$, while for $\MR=2\xi$, from equations \eqref{eq:MachISO} 
  and \eqref{eq:Wz_asymp} it can be shown that $\M^2 = 1+\Og(x)$.
  
\begin{figure}
        %\centering
        \includegraphics[width=1\linewidth]{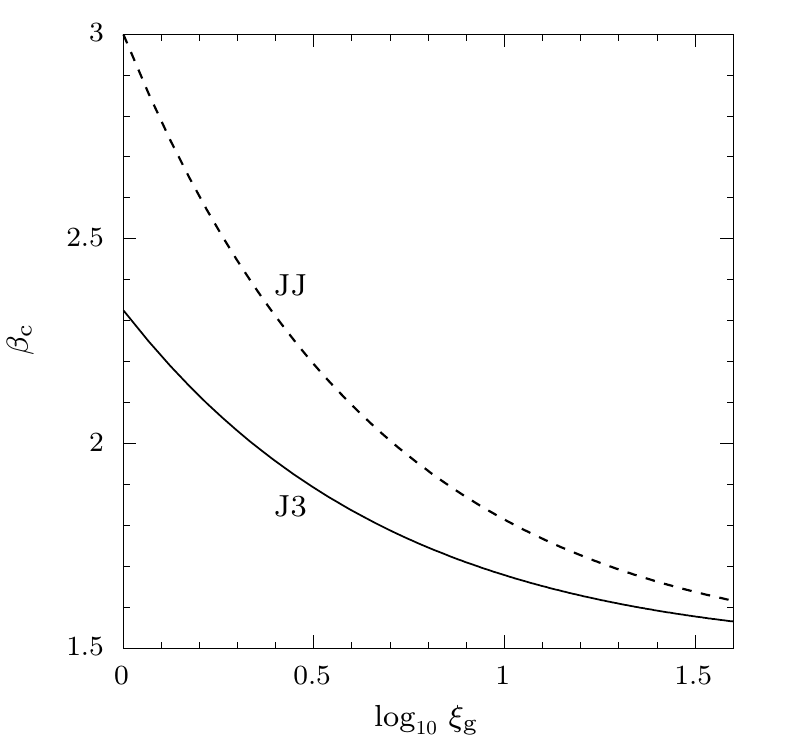}\\
        \vspace{-2.5mm}
        \caption{Critical temperature parameter 
                 $\bc \equiv 3/[\hspace{0.15mm}2\hspace{0.2mm}\Fg(\xig)]$
                 as a function of $\xig$, for J3 and JJ galaxy models. 
                 For $\beta=\Tinf/\TV>1$,
                 the isothermal accretion in absence of a central MBH is possible
                 provided that $\beta \leq \bc$. In these circumstances, once $\xig$ 
                 is fixed, the upper limit of $\beta$ for J3 models is lower
                 than that for JJ ones.}
  \label{fig:bc}
  \end{figure}

  We now show how the condition $\MR\geq 2\xi$ when $\chiup=0$, in order to have accretion,
  imposes an upper limit on $\Tinf$. In fact, from equation \eqref{eq:csib}, 
  with $\gamma=1$, the identity $\MR/(2\xi)=3/(2\beta\Fg)$ produces a condition
  for $\beta$\hspace{0.1mm}:
  \begin{equation}
        \beta \leq \frac{3}{2\Fg(\xig)} \equiv \bc\hspace{0.2mm},
  \label{eq:bc}
  \end{equation} 
  where the critical parameter $\bc$ depends only on $\xig$.
  It follows that {\it in absence of a central MBH, isothermal accretion in J3 galaxies 
  is possible only for}
  \begin{equation}
        \Tinf \leq \bc\hspace{0.3mm}\TV\hspace{0.2mm},
        \quad\,
        {\rm i.e.},
        \quad\,
        \sigpg(0)=\sigg(0)\geq\cinf\hspace{0.4mm},
  \end{equation}
  where the last inequality derives from equation \eqref{eq:R/xi}.
  For reference, in Fig. \ref{fig:bc} (right panel) we show  
  $\bc$ as a function of $\xig$, for both JJ and J3 models; it is easy to prove that
  $3/2 < \bc \leq 9/(\upi^2\!-6)\simeq 2.33$. 
  
  As anticipated, the limitation $\MR\geq 2\xi$ when $\chiup=0$ is 
  also relevant for the understanding of the flow behaviour when $\chiup>0$. 
  In fact, it is possible
  to show that, by defining $\tau \equiv \beta/\bc = 2\xi/\MR$, for $\MR\to\infty$  
  and fixed\footnote{As for JJ models (CP18, equation (48)), from equation \eqref{eq:xminISO} it follows
  that the limit for $\MR\to\infty$ is {\it not} uniform in $\tau$.} 
  $\tau$, we have
  \begin{equation}  
        \xmin \sim
        \begin{cases}
              \displaystyle -\hspace{0.4mm}\frac{\tau+\Wmu(-\hspace{0.4mm}\tau\hspace{0.2mm}\e^{-\hspace{0.2mm}\tau})}{2}\hspace{0.4mm}\MR\hspace{0.4mm}, 
              \hspace{0.75cm} \tau < 1\hspace{0.2mm},
              \\[10pt]
              \displaystyle \sqrt{\frac{\chiup\hspace{0.2mm}\MR}{2}}\hspace{0.4mm}, 
              \hspace{2.755cm} \tau = 1\hspace{0.2mm},
              \\[14pt]
              \displaystyle \frac{\chiup\hspace{0.2mm}\tau}{2\hspace{0.2mm}(\tau-1)}\hspace{0.4mm}, 
              \hspace{2.45cm} \tau > 1.
        \end{cases}
  \label{eq:xmin_asymp}
  \end{equation}
  The trend of $\xmin$ as a function of $\beta$ is shown by the black solid line 
  in Fig. \ref{fig:rmin} (top right panel), for a minimum halo model with $\xig=13$ 
  and $\mu=0.002$. For example, equation \eqref{eq:xmin_asymp} allows to explain the drop 
  at increasing $\beta$ when $\tau$
  switches from being less than unity to being larger than unity, with 
  $\xmin \simeq \chiup/2$ independently of $\tau$;
  the black square point at $\rmin \simeq 57.34\,\rB$ correspond to
  $\beta=\bc\simeq 1.65$, well approximated by the value $57.01\,\rB$
  obtained with the previous equation.
  Equation \eqref{eq:xmin_asymp} allows us to find the behaviour of $\lt$ for large 
  values of $\MR$ (at fixed $\beta$). For example, in the peculiar case $\beta=\bc$ (i.e. $\tau=1$),
  an asymptotic analysis shows that $\lt\sim\e^{3/2}\MR^2/4$; 
  for simplicity, we do not 
  report the expression of $\lt$ for $\beta\neq\bc$, which can, however, be easily calculated.
  As shown in Fig. \ref{fig:lt} (left panel), the presence of the galaxy makes $\lt$ 
  several orders of magnitude larger than without it.
  
\begin{figure*}
        %\centering
        \includegraphics[width=0.326\linewidth]{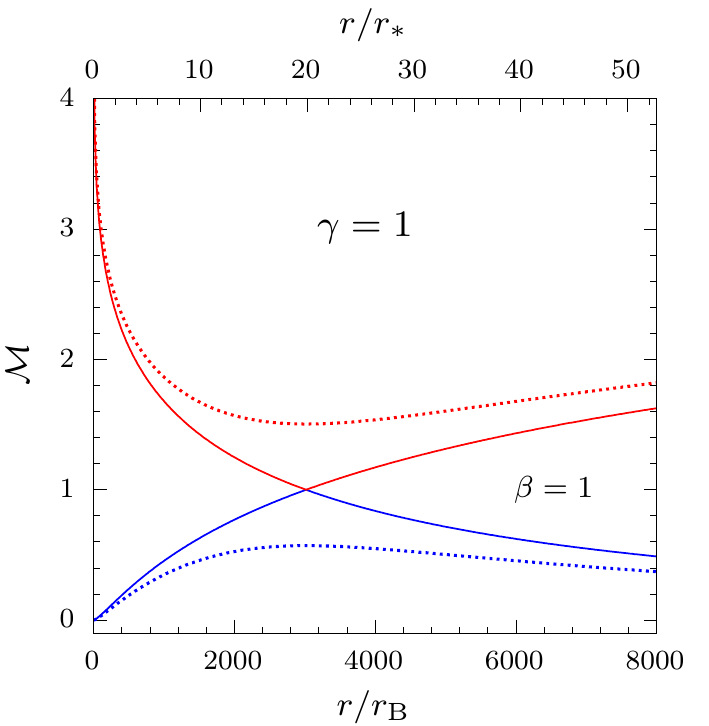}
        \,
        \includegraphics[width=0.326\linewidth]{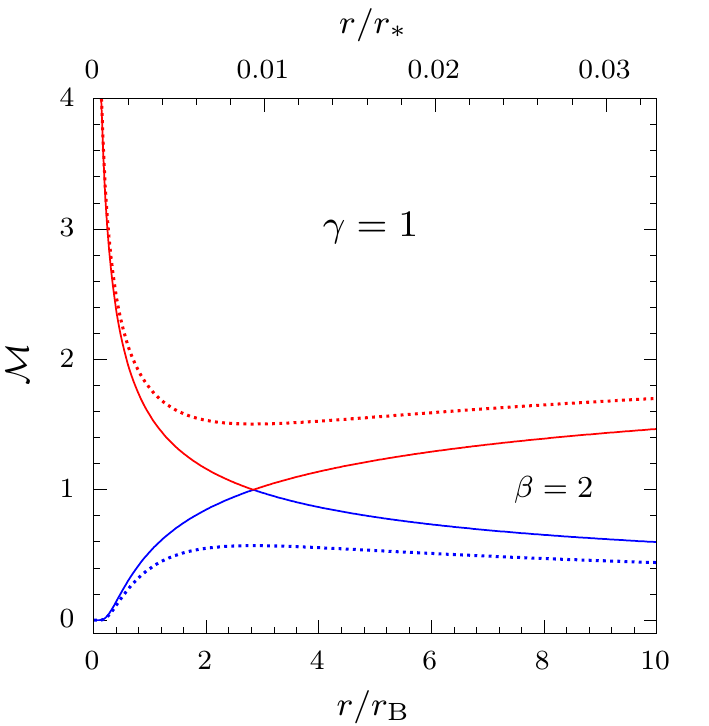}
        \,
        \includegraphics[width=0.326\linewidth]{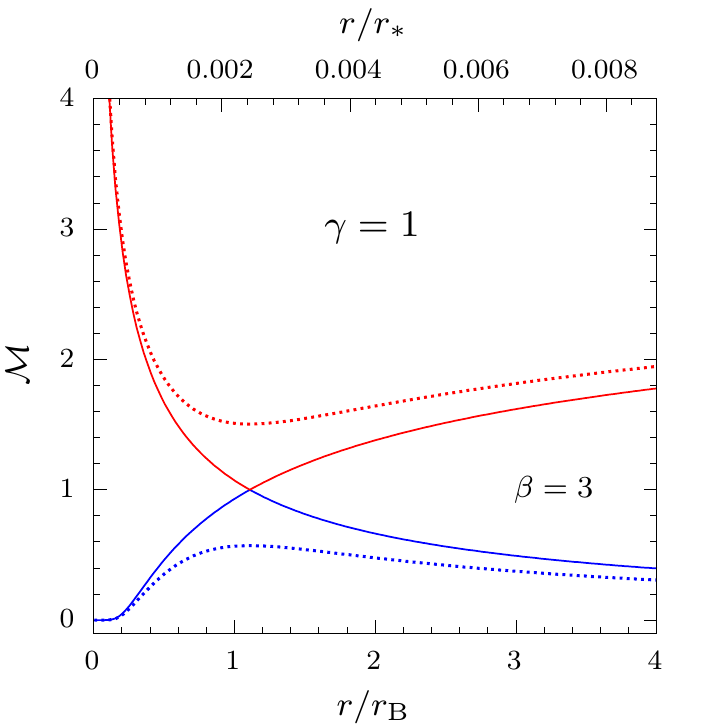}\\
        \vspace{4mm}
        \includegraphics[width=0.326\linewidth]{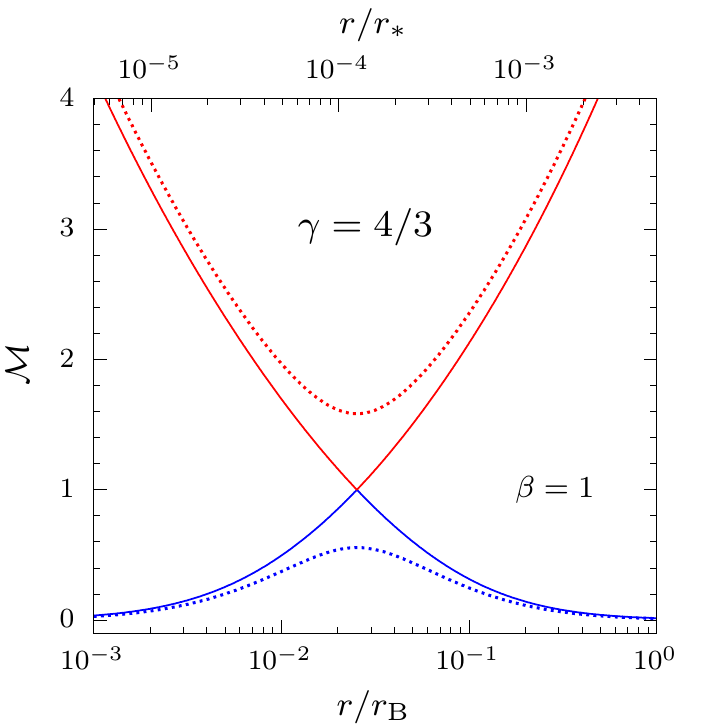}
        \,
        \includegraphics[width=0.326\linewidth]{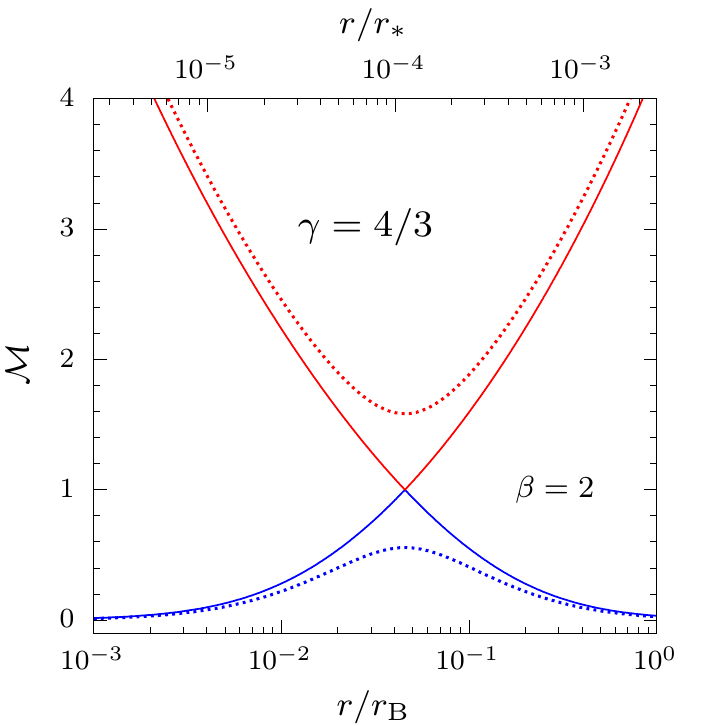}
        \,
        \includegraphics[width=0.326\linewidth]{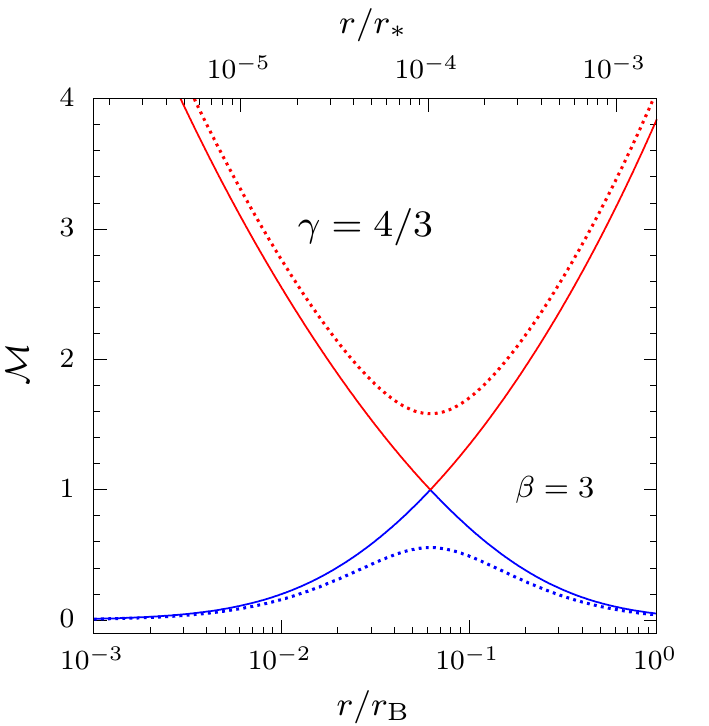}\\
        \vspace{4mm}
        \includegraphics[width=0.326\linewidth]{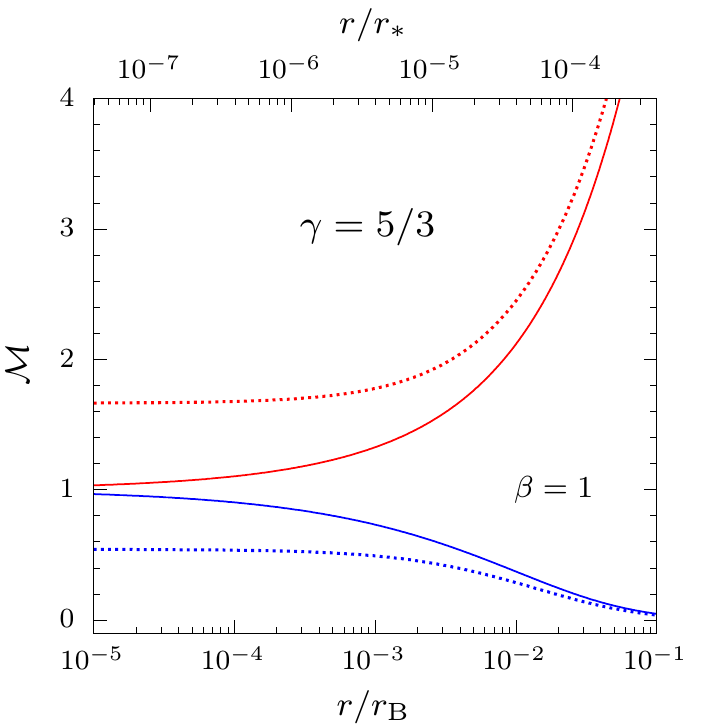}
        \,
        \includegraphics[width=0.326\linewidth]{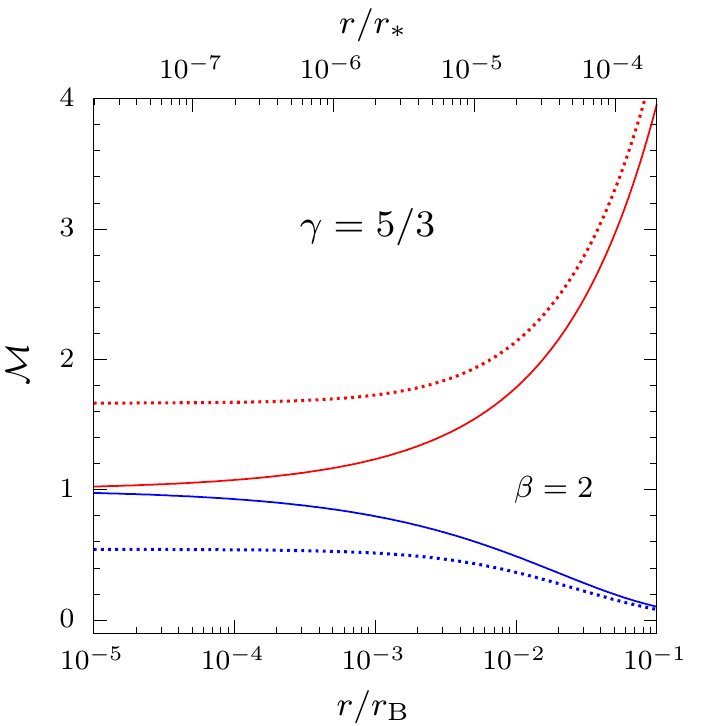}
        \,
        \includegraphics[width=0.326\linewidth]{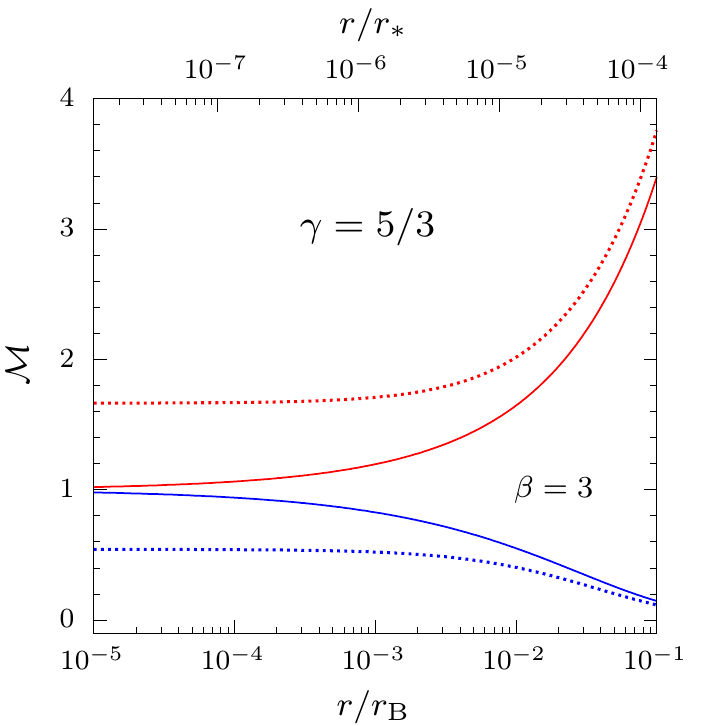}\\
        \caption{Radial profile of the Mach number for polytropic Bondi
        problem in a minimum halo J3 galaxy model with $\xig=13$, $\chiup=1$ and 
        $\mu=0.002$, in case of three different values of the gas temperature 
        ($\beta=1$, $2$, $3$). Solid lines show the two critical solutions ($\lambdaup=\lt$), 
        while dotted lines indicate the two subcritical solutions ($\lambdaup=0.8\,\lt$); 
        the distance from the centre is given in units of both $\rB$ 
        (bottom axis) and $\rs$ (top axis, using equation \eqref{eq:rB/rs}). 
        In blue we plot the subsonic regime and in red the supersonic one. 
        The top panels show the isothermal case ($\gamma=1$): notice how, in accordance with
        Fig. \ref{fig:rmin}, the position of $\rmin$ decreases very rapidly passing 
        from $\beta=1$ to $\beta=2$.
        Middle panels show the case $\gamma=4/3$: in accordance with
        the dashed black lines in Fig. \ref{fig:rmin}, $\rmin/\rs$ decreases for increasing
        $\beta$, while $\rmin/\rs$ decreases (note that a logarithmic scale for radius axes has
        been used).
        Finally, bottom panels show the adiabatic case ($\gamma=5/3$): the position of the sonic
        point is reached at the centre, the accretion solutions are always subsonic (i.e. $\M<1$), 
        and the wind solutions are always supersonic (i.e. $\M>1$).}
  \label{fig:Mach}
  \end{figure*}

  A summary of the results can be seen by inspection of 
  Fig. \ref{fig:Mach} (top panels), where we show the radial profile of the
  Mach number for three different values of the temperature parameter 
  ($\beta=1$, $2$, $3$). 
  Solid lines show the two critical solutions, one in which the gas flow begins supersonic
  and approaches the centre with zero velocity, and the other in which $\M$ continuously 
  increases towards the centre. 
  The dotted lines show two illustrative subcritical solutions with $\lambdaup=0.8\,\lt$.
  It is apparent that $\rmin$ decreases very rapidly with 
  increasing temperature at the transition from $\beta=1$ to $\beta=2$: 
  $\rmin \simeq 19.89\,\rs$, $0.0093\,\rs$, and $0.0024\,\rs$, for 
  $\beta=1$, $2$, and $3$, respectively.
  
\begin{figure}
        %\centering
        \includegraphics[width=1\linewidth]{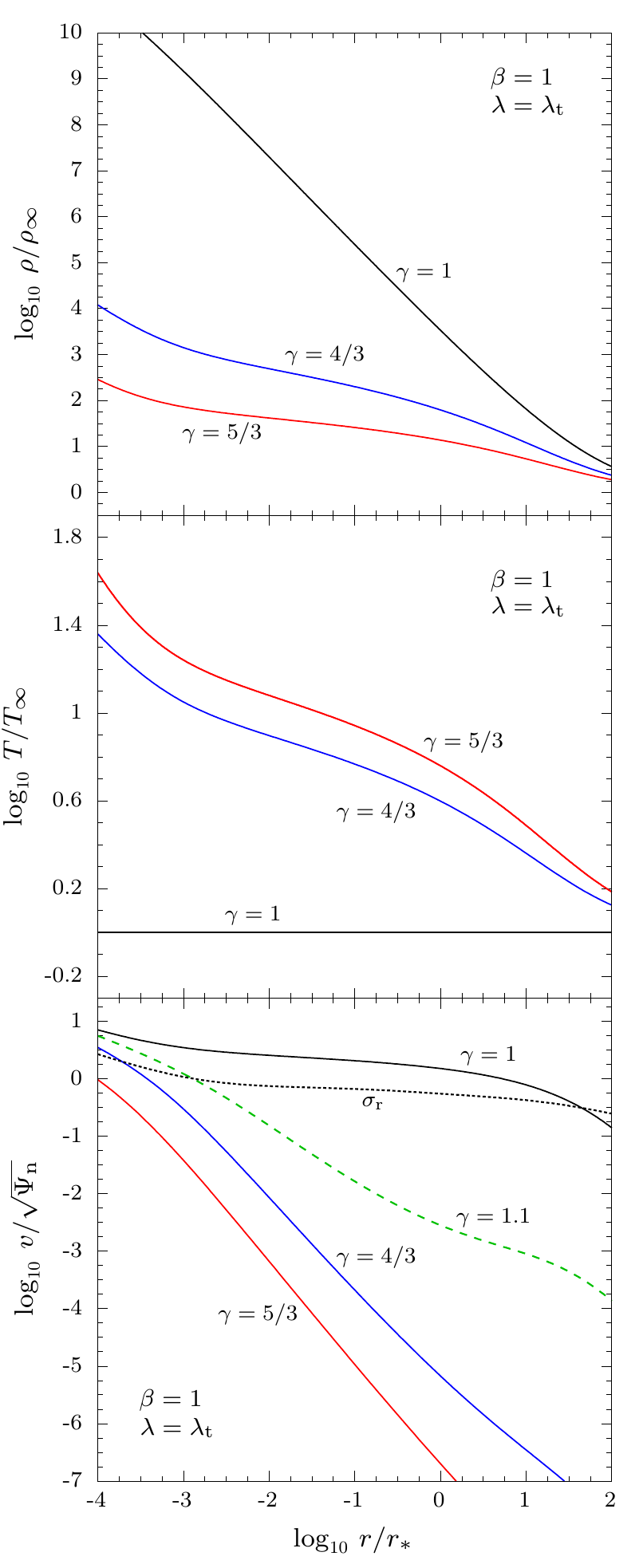}\\%0.331 se orizzontale
        \vspace{-3.5mm}
        \caption{Density (top), temperature (middle), and velocity (bottom) profiles, 
                 as a function 
                 of $r/\rs$, for the critical (i.e. $\lambdaup=\lt$) accretion 
                 solution of the polytropic 
                 Bondi problem in a minimum halo J3 galaxy model with $\xig=13$, 
                 $\chiup=1$, and $\mu=0.002$. The gas temperature at
                 infinity $\Tinf$ equals the stellar virial temperature $\TV$, i.e. 
                 $\beta=1$. For comparison, the dotted line in the bottom panel
                 shows the isotropic velocity dispersion profile $\sigr$.}
  \label{fig:rho(x)_T(x)}
  \end{figure}

  Finally, once the Mach number profile is known, the gas density profile
  is obtained from the first equation of the
  system \eqref{eq:cont+Bern_lambda} with $\gamma=1$, i.e.,
  \begin{equation}
        \rhotil(x)=\frac{\rho(x)}{\rhoinf}=\frac{\lambdaup}{x^2\M(x)}.
  \end{equation}
  Along the critical solution, by virtue of equation \eqref{eq:M2_asymp} it follows that 
  $\rhotil \sim \lt\hspace{0.2mm}x^{-\hspace{0.2mm}3/2}/\sqrt{2\chiup}$ at the centre when 
  $\chiup>0$, while $\rhotil \sim x^{\MR/x}$ at large radii.
  Fig. \ref{fig:rho(x)_T(x)} (top panel) shows the radial trend of $\rhotil$ 
  for the critical accretion solution in our reference model, with $\lt \simeq 2.14\times 10^8$. 
  The bottom panel shows the gas velocity profile and, for comparision,
  the isotropic velocity dispersion $\sigr$.
  Notice that near the centre, $\sigBH \propto r^{-\hspace{0.2mm}1/2}$ and 
  $\varv=\cinf\hspace{0.3mm}\M\propto r^{-\hspace{0.2mm}1/2}$ (provided that $\chiup>0$), 
  so that their ratio is constant; it can be easily shown that $\varv/\sigBH \sim 6\hspace{0.2mm}\chiup$.
  The value of $\sigBH$ near the centre (i.e. of $\sigr$ if a central MBH is present), 
  is then a proxy for the isothermal gas inflow velocity.
  %%%%%%%%%%%%%%%%%%%%%%%%%%%%%%%%%%%%%%%%%%%%%%%%%%%%%%%%%%%%%%%%%%%%%%%%%%%%%%
  %%%%%%%%%%%%%%%%%%%%%%%%%%%%%%%%%%%%%%%%%%%%%%%%%%%%%%%%%%%%%%%%%%%%%%%%%%%%%%
  
  %%%%%%%%%%%%%%%%%%%%%%%%%%%%%%%%%%%%%%%%%%%%%%%%%%%%%%%%%%%%%%%%%%%%%%%%%%%%%%
  %%%%%%%%%%%%%%%%%%%%%%%%%%%%%%%%%%%%%%%%%%%%%%%%%%%%%%%%%%%%%%%%%%%%%%%%%%%%%%
  %%%%%%%%%%%%%%%%%%%%%%%%%%%%%%%%%%%%%%%%%%%%%%%%%%%%%%%%%%%%%%%%%%%%%%%%%%%%%%
  %%%%%%%%%%%%%%%%%%%%%%%%%%%%%%%%%%%%%%%%%%%%%%%%%%%%%%%%%%%%%%%%%%%%%%%%%%%%%%
  
%%%%%%%%%%%%%%%%%%%%%%%%%%%%%%%%%%%%%%%%%%%%%%%%%%%%%%%%%%%%%%%%%%%%%%%%%  
  \subsection{The $1<\gamma<5/3$ case}
%%%%%%%%%%%%%%%%%%%%%%%%%%%%%%%%%%%%%%%%%%%%%%%%%%%%%%%%%%%%%%%%%%%%%%%%%
  
  When $1<\gamma<5/3$, from the expression for $f(x)$ we have that in one case the
  determination of $\xmin$ and $\fmin$ is trivial, i.e. for $\xi\to\infty$ 
  (or $\MR \to 0$)\hspace{0.1mm}: in this situation the galaxy contribution vanishes, 
  and the position of the only minimum of $f$ reduces to 
  $\xmin=\chiup\hspace{0.2mm}(5-3\gamma)/4$. Therefore, following 
  KCP16, the behaviour of the associated $\lt$ could be found just by carrying out
  a perturbative analysis (see KCP16, Appendix A); however, since in our models
  $\MR$ falls in the range $10^3 \div 10^4$, we shall not further discuss this limit case.
  In general, the problem of the determination of $\xmin$ (and so
  of $\lt$) cannot be solved analytically, as apparent by combining equations
  \eqref{eq:f_gen} and \eqref{eq:psi(x/csi)}, and setting $df/dx=0$\hspace{0.2mm}; 
  a numerical investigation is then needed.
  As in the case of isothermal flows, we begin by considering $0<\chiup\leq 1$.
  Of course, as $f$ is strictly positive, continuous, and divergent to infinity
  for $x \to 0$ and $x \to \infty$, the existence of at least a minimum is
  guaranteed. 
  A detailed numerical exploration shows that, in analogy with the isothermal
  case in Hernquist galaxies (CP17), it is possible to have more than one critical
  point for $f$ as a function of $\beta$ an $\gamma$.
  In particular, there can be a single minimum for $f$, or two minima and one maximum.
  We found that for $\xig=13$ and $\beta \approx 1 \div 2$, only one minimum is
  present for $\gamma \lesssim 1.01$ and $\gamma \gtrsim 1.1$; instead, 
  for $1.01 \lesssim \gamma \lesssim 1.1$, 
  three critical points and two minima are present.
  When $\beta$ is small (i.e. $\Tinf$ is low), the absolute minimum of $f$ is
  reached at the outer critical point; as $\beta$ increases, the value of $f$ 
  at the inner critical point decreases, and the flow is finally characterised
  by two sonic points. Increasing further $\Tinf$, the inner 
  critical point becomes the new sonic point, with a jump to a smaller value.
  Fig. \ref{fig:rmin} (top right panel) shows the position of $\xmin$ as a function 
  of $\beta$ for different values of $\gamma$, and confirms these trends of
  $\xmin$ with $\Tinf$ and $\gamma$. 
  Notice how the location of $\xmin$ 
  (shown in the bottom panel) now decreases with 
  an extremely slow decline for $\gamma>1$. 
  According with equation \eqref{eq:rminrs}, this means that, for polytropic indeces 
  sufficiently greater than $1$, the ratio $\rB/\rs$ decreases faster than what 
  $\xmin$ increases.

  In the case $\chiup=0$ the sonic point is reached 
  at the origin. Indeed, $f$ 
  tends to zero when $x \to 0$, and runs to infinity for $x \to \infty$, and so
  $\xmin \to 0$ for every choice of the model parameters. Therefore, from equation
  \eqref{eq:lambda_cr} one has $\lt \to 0$, concluding the discussion
  of the problem in absence of the central MBH since no accretion can take place.
  
  Having determined the position $\xmin$, we can compute 
  numerically the corresponding value of $\lt$, given in the polytropic case 
  by equation \eqref{eq:lambda_cr} with $\fmin=f(\xmin)$ obtained from equation \eqref{eq:f_gen}.
  In Fig. \ref{fig:lt} (right panel) the critical accretion parameter is shown as a function
  of $\xig$, for a reference model with $\gamma=4/3$ and different values of $\beta$.
  We note that, at variance with the isothermal case (left panel), $\lt$ is roughly constant 
  for fixed $\beta$ independently of the extension of the DM halo, while, at fixed $\xig$, it 
  increases for decreasing $\Tinf$. 
  Having also determined $\lt$, we finally solve numerically 
  equation \eqref{eq:Bondi_eq_poly}, obtaining the Mach profile $\M(x)$.
  In Fig. \ref{fig:Mach} (middle panels) we show $\M(x)$
  for three different values of the temperature parameter 
  ($\beta=1$, $2$, $3$).
  The logarithmic scale allows to appreciate how, according to Fig. \ref{fig:rmin},
  $\xmin$ suddenly falls down to values under unity as $\gamma$ increases with respect 
  to the isothermal case. 
  As an illustrative example, we show the case $\gamma=4/3$. 
  Although the trend is not very strong, the location of the sonic point, 
  at variance with the $\gamma=1$ case, moves away from the 
  centre as the temperature increases: $\rmin \simeq 0.025\,\rB$, $0.046\,\rB$, and $0.062\,\rB$,
  for $\beta=1$, $2$, and $3$, respectively. For comparison, in the top axis we give the
  distance from the origin in units of $\rs$, from which it can be seen that, in accordance
  with Fig. \ref{fig:rmin} (bottom panel), $\rmin$ now tends to increase slightly, 
  while still of the order of $10^{-4}\,\rs$.
  
  Once the radial profile of the Mach number is known, both the gas density and 
  temperature profiles can be obtained from the following relations:  
  \begin{equation}
        \rhotil\hspace{0.25mm}
        =\hspace{0.3mm}\tT^{\hspace{0.35mm}\frac{1}{\gamma-1}}\hspace{-0.2mm}
        =\hspace{0.3mm}\left(\frac{\lambdaup}{x^2\M}\right)^{\hspace{-0.2mm}\frac{2}{\gamma+1}},
  \label{eq:chain}
  \end{equation}
  with $\tT=T/\Tinf$.
  Fig. \ref{fig:rho(x)_T(x)} shows the trends of $\rho$ (top panel) and $T$ (middle panel),  
  as a function of $r/\rs$, for the critical accretion 
  solution in our usual reference model.
  The parameter $\beta$ is fixed to unity, and the curves refer to different 
  polytropic indeces.
  
  For what concerns the Mach profile for the critical accretion solution, 
  an asymptotic analysis of equation \eqref{eq:Bondi_eq_poly} shows that, at the leading order
  \begin{equation}
        \M \sim
        \begin{cases}
              \displaystyle{\hspace{0.2mm}\lt^{-\frac{\gamma-1}{2}}(2\hspace{0.1mm}\chiup)^{\frac{\gamma+1}{4}}\hspace{0.2mm}x^{-\frac{5-3\gamma}{4}}}\hspace{0.2mm},
              \hspace{9mm} x \to 0\hspace{0.4mm},
              \\[10pt]
              \displaystyle{\hspace{0.2mm}\lt\hspace{0.45mm} x^{-\hspace{0.35mm}2}}\hspace{0.2mm},
              \hspace{2.725cm} x \to \infty\hspace{0.1mm}.
        \end{cases}
  \label{eq:M_poly}
  \end{equation}
  Notice that all the information about the specific galaxy model in the two
  regions is contained in the parameter $\lt$.
  Equation \eqref{eq:M_poly} allows us to find the asymptotic behaviour at small and
  large radii of the most important quantities concerning the Bondi accretion.
  Close to the centre, for example, 
  $\rhotil \sim \lt\hspace{0.2mm}x^{-\hspace{0.2mm}3/2}/\sqrt{2\chiup}\hspace{0.3mm}$
  (as for the isothermal case),
  independently on the value of $\gamma$, and so for
  the gas velocity $\varv=\cs\hspace{0.3mm}\M$ one finds
  \begin{equation}
        \varv^2(r) 
        \sim \frac{\Psin\hspace{0.2mm}2\hspace{0.2mm}\chiup\hspace{0.2mm}\mu}{s} 
        \sim 6\hspace{0.2mm}\chiup\hspace{0.3mm}\sigBH^2(r)\hspace{0.2mm}.
  \end{equation}
  Therefore, the central value of $\sigBH$ is a proxy for the gas inflow velocity
  also in the range $1 < \gamma < 5/3$. 
  Fig. \ref{fig:rho(x)_T(x)} (bottom panel) shows the radial trend of $\varv$ for
  for different values of $\gamma$: notice how, moving away from the centre, it 
  decreases progressively faster for $\gamma>1$ (see the green dashed line, corresponding 
  to $\gamma=1.1$), while deviating significantly from the isotropic stellar velocity 
  dispersion profile.  
  
  We conclude by noting that the inclusion of the effects of the 
  gravitational field of an host galaxy allows to estimate the total mass profile,
  $\MT(r)=\MBH+\Mg(r)$, under the assumption of hydrostatic equilibrium
  (see e.g. Ciotti \& Pellegrini 2004; Pellegrini \& Ciotti 2006).
  First of all, we note that the estimated mass reads
  \begin{equation}
        \Mest(r)=
        \MT(r)+\frac{r^2}{2G}\frac{d\varv^2}{dr}\hspace{0.3mm}, 
  \end{equation}
  whence it is clear that the hypothesis of hydrostatic equilibrium always leads to underestimate $\MT$ 
  in the accretion studies, where the velocity increases in magnitude towards the centre.
  Simple algebra shows that the expression of $\Mest$ is given by
  \begin{equation}
        \Mest(r)
        =-\hspace{0.4mm}\frac{r^2}{G\rho(r)}\hspace{0.1mm}\frac{dp}{dr}
        =-\hspace{0.4mm}\MBH\hspace{0.4mm}\frac{x^2}{\rhotil^{\hspace{0.3mm}2-\hspace{0.3mm}\gamma}}\hspace{0.1mm}\frac{d\rhotil}{dx},
  \end{equation}
  and, near the MBH (i.e. for $x \to 0$), where 
  $\rhotil \sim \lt\hspace{0.2mm}x^{-3/2}/\sqrt{2\chiup}\hspace{0.3mm}$, 
  \begin{equation}
        \frac{\Mest(r)}{\MBH} \sim 
        \frac{3}{2}\left(\frac{\lt}{\sqrt{\hspace{0.2mm}2\hspace{0.1mm}\chiup\hspace{0.3mm}}}\hspace{0.3mm}\right)^{\gamma-1}
        \hspace{-0.2mm}x^{\frac{5-3\gamma}{2}};
  \label{eq:Mest_poly}
  \end{equation}    
  notice that in the isothermal limit case one has $\Mest(r) \propto r$.
  
%%%%%%%%%%%%%%%%%%%%%%%%%%%%%%%%%%%%%%%%%%%%%%%%%%%%%%%%%%%%%%%%%%%%%%%%%  
  \subsection{The $\gamma=5/3$ case}
%%%%%%%%%%%%%%%%%%%%%%%%%%%%%%%%%%%%%%%%%%%%%%%%%%%%%%%%%%%%%%%%%%%%%%%%%

  The monoatomic case ($\gamma=5/3$) presents some special behaviour 
  deserving a short description.
  By considering equation \eqref{eq:f_gen} with $\gamma=5/3$, it follows that
  $f$ is monotonically increasing and the only 
  minimum is reached at the centre (KCP17); moreover, for galaxy models with 
  $r\hspace{0.1mm}\Psig(r) \to 0$ when $r \to 0$ (as for J3 models), one
  finds $\fmin=\chiup$, whence $\lt=\chiup^2/4$. 
  Therefore, $\chiup>0$ in order to have accretion.
  
  When $\lambdaup=\chiup^2/4$, the Bondi problem \eqref{eq:Bondi_eq_poly} reduces 
  to the fourth degree equation
  \begin{equation}
        \M^2\hspace{-0.2mm}-\frac{4f(x)}{\chiup}\hspace{0.35mm}\sqrt{\M\hspace{0.5mm}}+3=0\hspace{0.4mm},
  \end{equation}
  provided that the condition on the central potential mentioned
  above is satisfied; note that the dependence on the specific galaxy model is contained 
  only in the function $f(x)$.
  In the bottom panels of Fig. \ref{fig:Mach} we show the radial profile of the
  Mach number.
  In this situation, $\xmin=0$, and so the accretion solutions (blue lines) are 
  subsonic everywhere.  
  
  The asymptotic bahaviour of $\M(x)$ for the critical accretion solution when $x \to \infty$ is obtaned from 
  equation \eqref{eq:M_poly} just by fixing $\lt=\chiup^2\hspace{-0.2mm}/\hspace{0.2mm}4$.
  When $x \to 0$, instead, the $\gamma=5/3$ case does {\it not} coincide with the limit of 
  equation \eqref{eq:M_poly} for $\gamma \to 5/3$\hspace{0.1mm}: 
  in fact, now $\M \to 1$ instead of infinity, and its asymptotic trend reads
  \begin{equation}
        \M(x) \sim 1-\hspace{0.1mm}\sqrt{\hspace{0.4mm}-\hspace{0.4mm}\frac{8\hspace{0.1mm}\MR\hspace{0.3mm}x\ln x}{3\hspace{0.2mm}\xi\hspace{0.15mm}\chiup}\hspace{0.4mm}}\hspace{0.3mm},
        \qquad
        x \to 0\hspace{0.15mm};
  \end{equation}
  of course, the same situation at small radii occurs in the case of any 
  other quantity deriving from Mach's profile: for example,
  \begin{equation}
        \varv^2(r) \sim \frac{\Psin\hspace{0.2mm}\chiup\hspace{0.2mm}\mu}{2\hspace{0.1mm}s}\hspace{0.2mm},
        \qquad
        \Mest(r) \sim \frac{3\hspace{0.2mm}\chiup}{4}\hspace{0.3mm}\MBH\hspace{0.1mm}.
  \end{equation}  
  Notice that $\varv$ decreases by a factor of $2$ with respect to the $1 \leq \gamma < 5/3$ case,
  and $\Mest$ differs from what would be obtained 
  setting $\gamma=5/3$ and $\lt=\chiup^2\hspace{-0.2mm}/\hspace{0.2mm}4$ in 
  equation \eqref{eq:Mest_poly}. 
  
%%%%%%%%%%%%%%%%%%%%%%%%%%%%%%%%%%%%%%%%%%%%%%%%%%%%%%%%%%%%%%%%%%%%%%%%%
  \section{Entropy and Heat Balance Along The Bondi Solution}\label{sec:Entropy}
%%%%%%%%%%%%%%%%%%%%%%%%%%%%%%%%%%%%%%%%%%%%%%%%%%%%%%%%%%%%%%%%%%%%%%%%%  

  %%%%%%%%%%%%%%%%%%%%%%%%%%%%%%%%%%%%%%%%%%%%%%%%%%%%%%%%%%%%%%%%%%%%%%%%%%%%%%
  \begin{figure*}
        %\centering
        \includegraphics[width=0.48\linewidth]{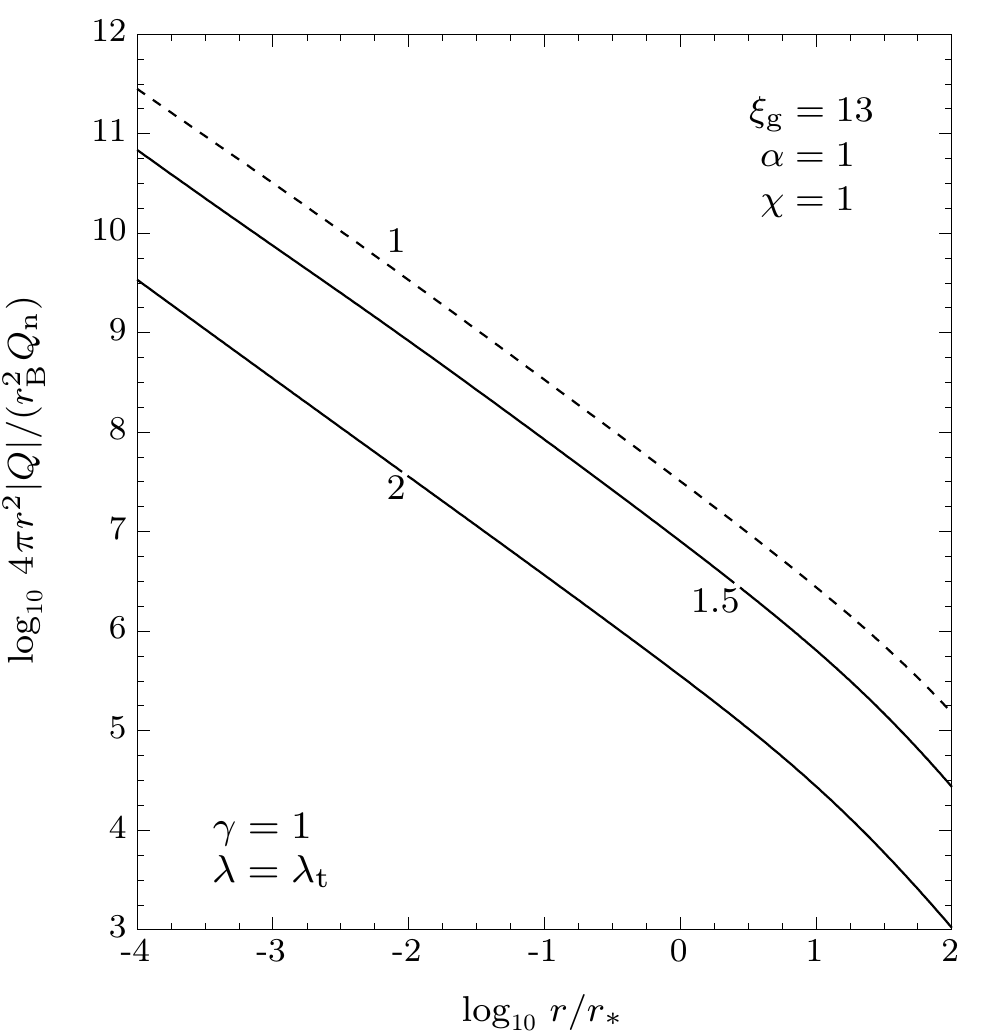}
        \quad\,\,\,
        \includegraphics[width=0.48\linewidth]{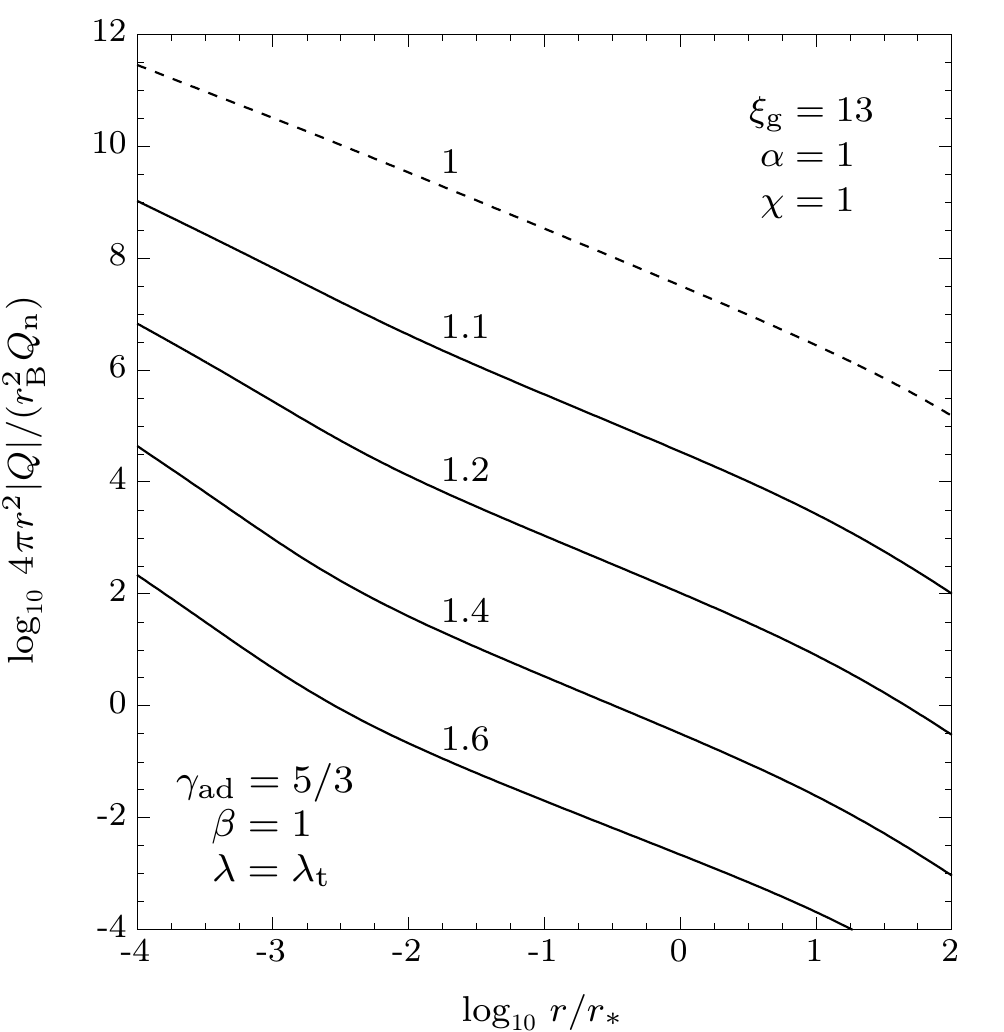}\\
        %\vspace{1mm}
        \caption{Absolute value of the rate of heat per unit lenght exchanged by the fluid element, 
                 $4\hspace{0.1mm}\upi r^2|\Qdot\hspace{0.15mm}|$, in units 
                 of $\rB^2\hspace{0.1mm}\Qn$, 
                 as a function $r/\rs$, for the critical Bondi accretion of 
                 a minimum halo J3 model with $\xig=13$, $\chiup=1$, and $\mu=0.002$.   
                 Left panel: isothermal case for $\beta=1$, $1.5$, and $2$.
                 Right panel: monoatomic gas 
                 ($\gammaAD=5/3$) with $\beta=1$, for different values of the polytropic index.
                 In both panels, the dashed lines correspond to isothermal
                 accretion with $\Tinf=\TV$.}
  \label{fig:4pir2Q}
  \end{figure*}
  %%%%%%%%%%%%%%%%%%%%%%%%%%%%%%%%%%%%%%%%%%%%%%%%%%%%%%%%%%%%%%%%%%%%%%%%%%%%%
  
  In this Section we employ the obtained polytropic solutions to elucidate some
  important thermodynamical aspects of the Bondi accretion, not always sufficiently
  stressed in the literature. In fact, it is not uncommon to consider Bondi accretion
  as an `adiabatic' problem, where no radiative losses or other forms of heat 
  transfer take place: after all, no heating or cooling functions seem to be specified 
  at the outset of the problem. Obviously, this is not true, being 
  the Bondi solution a purely hydrodynamical flow where all the 
  thermodynamics of heat exchange is implicitly described by the polytropic
  index $\gamma$. 
  Therefore, for given $\gamma$ (and in absence of shock
  waves), one can follow the entropy evolution of each fluid element along the 
  radial streamline, and determine the reversible heat exchanges.
  Let us consider polytropic Bondi accretion with\footnote{Notice that not necessarily 
  $\gamma<\gammaAD$; for example, one could study
  a $\gamma=5/3$ accretion in a biatomic gas with $\gammaAD=7/5$.} 
  $\gamma\neq\gammaAD$.   
  From the expression of the entropy per unit mass $\s$ for a perfect gas
  (e.g. Chandrasekhar 1939; Zel'dovich \& Raizer 1966), and assuming
  as reference value for $\s$ its value at infinity, 
  we can write the change of entropy of an element of the
  accreting flow along its radial streamline, during a polytropic transformation, as
  \begin{equation}
  \frac{D\s}{Dt}=\cV\hspace{0.2mm}(\hspace{0.15mm}\gamma-\gammaAD)\hspace{0.4mm}
  \frac{D\ln\rhotil}{Dt}\hspace{0.2mm},
  \qquad\,\,
  \Delta\s \equiv \s-\hspace{0.15mm}\sinf\hspace{0.2mm},
  \label{eq:Ds/Dt}
  \end{equation}
  where $D/Dt=\partial/\partial t+{\bm \varv}\hspace{0.15mm}\cdot\Grad$ is 
  the material derivative.
  Of course, for $\gamma=\gammaAD$ no change of entropy occurs along regular solutions,
  being the process isentropic; instead, for $\gamma\neq\gammaAD$, once the 
  solution of the Bondi problem is known, equation \eqref{eq:Ds/Dt} allows
  to compute the entropy change of a fluid element. 
  From the second law of thermodynamics, the rate of heat per unit mass 
  exchanged by the fluid element can be written as 
  \begin{equation}
        \frac{Dq}{Dt}=T\hspace{0.25mm}\frac{D\s}{Dt}\hspace{0.1mm}.
  \label{eq:DQ/Dt_s}
  \end{equation}   
  Therefore, from equation \eqref{eq:Ds/Dt}, it follows that, for $\gamma\neq\gammaAD$, a fluid element
  necessarily exchanges heat with the ambient; this fact can be restated in terms of
  the specific heat as
  \begin{equation}
        \frac{Dq}{Dt}=\mathcal{C}\hspace{0.5mm}\frac{DT}{Dt}
                     =\cV\hspace{0.2mm}\frac{\gamma-\gammaAD}{\gamma-1}\hspace{0.1mm}\frac{DT}{Dt}
                     \hspace{0.1mm},
  \label{eq:DQ/Dt_T}  
  \end{equation}
  where $\mathcal{C}$ is the {\it constant} specific heat for polytropic trasformations
  (see e.g. Chandrasekhar 1939).
  A third (equivalent) expression for the heat exchange can be
  finally obtained from the first law of thermodynamics, i.e., 
  \begin{equation}
        \frac{Dq}{Dt}
        =\frac{De}{Dt}-\frac{p}{\rho^2}\frac{D\rho}{Dt}\hspace{0.2mm},
  \label{eq:DQ/Dt_First}
  \end{equation}
  where $e$ is the internal energy per unit mass, and, 
  apart from an additive constant,
  $h=e+p/\rho=\cp\hspace{0.15mm}T$ is the enthalpy per unit mass.
  In the stationary case, from equations \eqref{eq:DQ/Dt_s}, \eqref{eq:DQ/Dt_T}
  and \eqref{eq:DQ/Dt_First}, one has 
  \begin{equation}
        \frac{\Qdot}{\rho}\equiv\frac{Dq}{Dt}=
        \begin{cases}
        \hspace{0.5mm}\displaystyle\cV\hspace{0.2mm}(\hspace{0.15mm}\gamma-\gammaAD)\hspace{0.5mm}T\hspace{0.3mm}{\bm \varv}\hspace{0.15mm}\cdot\frac{\Grad\rho}{\rho}\hspace{0.2mm},
        \\[11pt]
        \hspace{0.5mm}\displaystyle \cV\hspace{0.2mm}\frac{\gamma-\gammaAD}{\gamma-1}\hspace{0.6mm}{\bm \varv}\hspace{0.15mm}\cdot\Grad\hspace{0.25mm}T
                     \hspace{0.1mm},
        \\[10pt]
        \hspace{0.5mm}\displaystyle {\bm \varv}\hspace{0.15mm}\cdot\Grad\hspace{-0.25mm}
        \left(\frac{\varv^2}{2}+h-\PsiT\hspace{-0.2mm}\right)\hspace{-0.25mm},
        \end{cases}
  \label{eq:DQ/Dt}
  \end{equation} 
  where $\Qdot$ is the rate of heat exchange per unit volume, 
  ${\bm \varv}=-\hspace{0.5mm}\varv\hspace{0.25mm}\er$, 
  $\Grad=\er\hspace{0.2mm}d/dr$, and the last expression can be easily proved 
  (e.g. Ciotti 2021, Chapter 10).
  Summarising, a fluid element undergoing a generic polytropic transformation
  loses energy as it moves inward and heats when $1<\gamma<\gammaAD$, while
  for $\gamma>\gammaAD$ it experiences a temperature decrease. 
  In the polytropic Bondi accretion both cases are possible, 
  except for a monoatomic gas, when accretion is possible {\it only} for 
  $\gamma \leq \gammaAD = 5/3$ (see Section \ref{sec:Bondi}). 
  We can now use each expression in equation \eqref{eq:DQ/Dt} to compute the rate 
  of heat exchange just by substituting in them the solution of the Bondi problem.
  Defining $\Qn=\cinf^3\hspace{0.2mm}\rhoinf/\rB$, the first two expression in 
  \eqref{eq:DQ/Dt}, and the third one, become respectively
  \begin{equation}
        \Qdot=\frac{\Qn\hspace{0.2mm}\lt}{x^2}\times
        \begin{cases}
          \hspace{0.2mm}\displaystyle\frac{\gammaAD-\hspace{0.2mm}\gamma}{\gamma\hspace{0.2mm}(\gammaAD-\hspace{0.2mm}1)}\hspace{0.4mm}
        \rhotil^{\hspace{0.5mm}\gamma-2}\hspace{0.4mm}\frac{d\rhotil}{dx}\hspace{0.2mm},
        \\[14pt]
          \hspace{0.2mm}\displaystyle-\hspace{0.4mm}\frac{d\E}{dx}\hspace{0.2mm},
        \end{cases}
  \label{eq:Q_normal}
  \end{equation}
  where, up to an additive constant,
  \begin{equation}
        \E \equiv 
        \left[\hspace{0.1mm}\frac{\M^2}{2}+\frac{\gammaAD}{\gamma(\gammaAD-1)}\hspace{0.1mm}\right]\hspace{-0.6mm}\rhotil^{\hspace{0.5mm}\gamma-1}
        -\hspace{0.2mm}\left[\hspace{0.3mm}\frac{\chiup}{x}+\frac{\MR}{\xi}\hspace{0.4mm}\psi\!\left(\frac{x}{\xi}\right)\hspace{-0.2mm}\right]\hspace{-0.4mm}.
  \end{equation}
  The situation is illustrated in Fig. \ref{fig:4pir2Q}: the left panel refers
  to the isothermal case and three values of $\beta$; the right panel
  shows the case of a {\it monoatomic} gas (i.e. $\gammaAD=5/3$), for 
  a fixed $\beta$ and different values of $\gamma<\gammaAD$.
  The plotted quantity is 
  $-\hspace{0.5mm}4\hspace{0.1mm}\upi\hspace{0.1mm}r^2\Qdot(r)$, i.e. the rate 
  of heat per unit lenght exchanged by the infalling gas element.
  In practice, by integrating the curves between two radii $r_1$ and $r_2$, one  
  obtain the heat per unit time exchanged with the ambient by the spherical shell of 
  thickness $|\hspace{0.3mm}r_2-r_1|$.
  For comparison, the dashed lines correspond to the same case, i.e. 
  isothermal accretion with $\Tinf=\TV$. Notice how in general the profile is almost
  a power law over a very large radial range, and how the heat exchange decreases for
  increasing $\Tinf$ and for $\gamma$ approaching $\gammaAD$.
  
  An important region for observational and theoretical works is the galactic centre.
  The general asymptotic trend of $\Qdot$, for $x \to 0$ and $\chiup>0$, reads 
  \begin{equation}
        \frac{\Qdot}{\Qn}\sim
        \begin{cases}
          \hspace{0.2mm}\displaystyle\frac{3\hspace{0.2mm}\lt^{\gamma}\hspace{0.2mm}(2\hspace{0.1mm}\chiup)^{-\frac{\gamma-1}{2}}(\gamma-\hspace{0.2mm}\gammaAD)}{2\hspace{0.2mm}\gamma\hspace{0.2mm}(\gammaAD-\hspace{0.2mm}1)}\hspace{0.5mm}x^{-\frac{3(\gamma+1)}{2}}
        \sim \frac{3\hspace{0.1mm}\chiup\hspace{0.3mm}(\gamma-\hspace{0.2mm}\gammaAD)}{\lt\hspace{0.2mm}\gamma\hspace{0.2mm}(\gammaAD-\hspace{0.2mm}1)}\hspace{0.5mm}\rhotil^{\hspace{0.4mm}2}\hspace{0.2mm}\tT,
        \\[14pt]
          \hspace{0.2mm}\displaystyle\frac{3\hspace{0.2mm}\chiup^3(5-3\gammaAD)}{80\hspace{0.3mm}(\gammaAD-\hspace{0.2mm}1)}\hspace{0.3mm}x^{-\hspace{0.4mm}4}
        \sim \frac{3\hspace{0.3mm}(5-3\gammaAD)}{5\hspace{0.2mm}\chiup\hspace{0.3mm}(\gammaAD-\hspace{0.2mm}1)}\hspace{0.5mm}\rhotil^{\hspace{0.4mm}2}\hspace{0.2mm}\tT,
        \end{cases}
  \end{equation} 
  where in the first expression, $1\leq\gamma< 5/3$ and
  $\rhotil \sim \lt\hspace{0.2mm}x^{-\hspace{0.2mm}3/2}/\sqrt{2\chiup}\hspace{0.2mm}$, 
  and in the second, $\gamma=5/3$ and $\rhotil \sim (\chiup/2)^{3/2}x^{-\hspace{0.2mm}3/2}$.
  In practice, close to the centre, $\Qdot$ is a pure power law of logarithmic slope 
  decreasing from $-\hspace{0.5mm}3$ to $-\hspace{0.6mm}4$ for $\gamma$  
  increasing from $1$ to $5/3$.
  It follows that the volume integrated heat exchanges are always dominated by
  the innermost region.
  
  We conclude by noticing the interesting fact that the heat per unit mass exchanged by
  a fluid element as it moves from $\infty$ down to the radius $r$, admits a very simple
  physical interpretation; in fact, by integrating the last expression of
  equation \eqref{eq:DQ/Dt} along the streamline, 
  one obtains for this exchange the remarkable result that
  \begin{equation}
        \Delta q=\frac{\varv^2}{2}+\Delta h-\PsiT\hspace{0.3mm},
        \qquad
        \Delta h \equiv h(r)-h(\infty)\hspace{0.2mm};
  \label{eq:Deltaq}
  \end{equation}
  the total heat exchanged by a unit mass
  of fluid (moving from $\infty$ to $r$) can then be interpreted as the change 
  of the Bernoulli `constant' when 
  the enthalpy change in equation \eqref{eq:Deltah-Deltaq} is evaluated along the 
  polytropic solution.
  There is an interesting alternative way to obtain the result above.
  In fact, from the first law of thermodynamics, $dq=dh-dp/\rho$, thus 
  in our problem we also have
  \begin{equation}
        \int_{\pinf}^p\frac{dp}{\rho}
        \hspace{0.25mm}=\hspace{0.25mm}\Delta h\hspace{0.2mm}-\hspace{0.2mm}\Delta q
        =\left(1-\frac{\mathcal{C}}{\cp}\right)\hspace{-0.3mm}\Delta h\hspace{0.2mm}.
  \label{eq:Deltah-Deltaq}
  \end{equation}
  This shows that the integral at the left hand side, which appears in Bondi accretion
  through equation \eqref{eq:Bernoulli},
  equals $\Delta h$ {\it only} for $\gamma=\gammaAD$, while, in general, 
  it is just {\it proportional} to $\Delta h$.
  Equation \eqref{eq:Deltaq} can also be obtained
  by inserting equation \eqref{eq:Deltah-Deltaq} in equation \eqref{eq:Bernoulli}, and
  considering the total potential (galaxy plus MBH).   

%%%%%%%%%%%%%%%%%%%%%%%%%%%%%%%%%%%%%%%%%%%%%%%%%%%%%%%%%%%%%%%%%%%%%%
  \section{Discussion and conclusions}\label{sec:Conclusions}
%%%%%%%%%%%%%%%%%%%%%%%%%%%%%%%%%%%%%%%%%%%%%%%%%%%%%%%%%%%%%%%%%%%%%%

  A recent paper (CP18) generalised the Bondi accretion theory 
  to include the effects of the gravitational field of the 
  galaxy hosting a central MBH, and of electron scattering,
  finding the analytical isothermal accretion solution for 
  Jaffe's two-component JJ galaxy models (CZ18).
  The JJ models are interesting because
  almost all their relevant dynamical properties can be expressed in 
  relatively simple analytical form, while reproducing the main structural
  properties of real ellipticals.
  However, their DM haloes cannot reproduce the expected $r^{-3}$ profile 
  at large radii, characteristic of the NFW profile; as Bondi accretion
  solution is determined by the gas properties at `infinity', it is important 
  to understand the effect of a more realistic DM potential at large radii.
  Moreover, in CP18 only isothermal solution were studied.
  Later, CMP19 presented two-component J3 galaxy models, similar to the JJ ones
  but with the additional property that the DM halo can reproduce the NFW profile 
  at all radii. J3 models then represent an improvement over JJ ones, while 
  retaining the same analytical simplicity, and so avoiding the need for numerical investigations to 
  study their dynamical properties. 
  In this paper we take advantage of J3 models to study again the generalised Bondi problem, 
  further extending the investigation to the general case of a polytropic gas,
  and elucidating some important thermodynamical properties of accretion. 
  The parameters describing the solution are linked to the galaxy structure 
  by imposing that the gas temperature at infinity ($\Tinf$) is proportional 
  to the virial temperature of the stellar component ($\TV$) through a dimensionless
  parameter ($\beta$) that can be arbitrarily fixed. 
  The main results can be summarised as follows.
  \begin{enumerate}
        \item The isothermal case can be solved in a fully analytical way. 
              In particular, there is only one sonic point for any choice of the
              galaxy structural parameters and of the value of $\Tinf$.
              It is found however that $\rmin$, the position of the sonic radius, is strongly
              dependent on $\Tinf$, with values of the order of, or larger than, the 
              galaxy effective radius ($\Reff$) for temperatures of the order of
              $\TV$, and with a sudden decrease down to $\approx 10^{-2}\hspace{0.3mm}\Reff$,
              or even lower, at increasing $\Tinf$ (say $\gtrsim 1.5\hspace{0.6mm}\TV$). 
              In absence of a central MBH (or $\chiup=0$, i.e. when the gravitational
              attraction of the central MBH is perfectly balanced by the radiation pressure),
              accretion is possible provided that $\cinf\leq\sigpg(0)$, i.e. when $\Tinf$ is lower than a 
              critical value, with $\sigpg(0)$ the central projected stellar velocity dispersion. 
        \item When $1<\gamma<5/3$, the Bondi accretion problem does not allow 
              for an analytical solution.   
              A numerical exploration shows that $\rmin$ suddenly drops to 
              values $\lesssim\rs$ as $\gamma$ increases at fixed $\Tinf$.
              Moreover, depending on the specific values of $\MR$, 
              $\xi$, and $\gamma$, the accretion flow can have one or three critical points,
              and in very special circumstances two sonic points.
              For a given $\gamma$, quite independently of the extension of the DM halo, 
              the accretion parameter $\lt$ is roughly constant at fixed $\beta$, with values
              several order of magnitudes lower than the isothermal case. 
              In absence of a central MBH, no accretion can take place.
        \item In the monoatomic adiabatic case ($\gamma=5/3$) 
              the Mach number profile can be obtained for a generic galaxy model
              by solving a fourth degree algebraic equation. 
              However, the solution is quite impractical, and
              a numerical evaluation is preferred. 
              As already shown in KCP16, in this case $\lt=\chiup^2\hspace{-0.3mm}/\hspace{0.2mm}4$, 
              so that, again, the absence of the central MBH makes accretion 
              impossible.
        \item We consider in detail the
              thermodynamical properties of Bondi accretion when the polytropic index
              $\gamma$ differs from the adiabatic index $\gammaAD$.
              Under this circumstance, the entropy of fluid elements changes along their
              pathlines, and it is possible to compute the associated heat exchanges ($\Qdot$\hspace{0.3mm}).
              We provide the mathematical expressions to compute $\Qdot$ as a function
              of radius, once the Bondi problem is solved, and in particular its 
              asymptotic behaviour near the MBH.
  \end{enumerate}
              
%%%%%%%%%%%%%%%%%%%%%%%%%%%%%%%%%%%%%%%%%%%%%%%%%%%%%%%%%%%%%%%%%%%%%%%%%%%%%%%%
%  \section*{Acknowledgements}
%%%%%%%%%%%%%%%%%%%%%%%%%%%%%%%%%%%%%%%%%%%%%%%%%%%%%%%%%%%%%%%%%%%%%%%%%%%%%%%%

%  We thanks... \textcolor{red}{\bf To Be Written?}
  
%%%%%%%%%%%%%%%%%%%%%%%%%%%%%%%%%%%%%%%%%%%%%%%%%%%%%%%%%%%%%%%%%%%%%%%%%%%%%%%%
  \section*{Data Availability}
%%%%%%%%%%%%%%%%%%%%%%%%%%%%%%%%%%%%%%%%%%%%%%%%%%%%%%%%%%%%%%%%%%%%%%%%%%%%%%%%

  No datasets were generated or analysed in support of this research.

%%%%%%%%%%%%%%%%%%%%%%%%%%%%%%%%%%

%%%%%%%%%%%%%%%%%%%%%%%%%%%%%%%%%%%%%%%%%%%%%%%%%%%%%%%%%%%%%%%%%%%%%%%%%%%%%%%%%%%%%%%%%%%%%%
%***************************************APPENDICI*********************************************
%%%%%%%%%%%%%%%%%%%%%%%%%%%%%%%%%%%%%%%%%%%%%%%%%%%%%%%%%%%%%%%%%%%%%%%%%%%%%%%%%%%%%%%%%%%%%%

  \appendix
%%%%%%%%%%%%%%%%%%%%%%%%%%%%%%%%%%%%%%%%%%%%%%%%%%%%%%%%%%%%%%%%%%%
  \section{The Lambert\hspace{0.2mm}-\hspace{0.4mm}Euler function}\label{app:W}
%%%%%%%%%%%%%%%%%%%%%%%%%%%%%%%%%%%%%%%%%%%%%%%%%%%%%%%%%%%%%%%%%%%

  %%%%%%%%%%%%%%%%%%%%%%%%%%%%%%%%%%%%%%%%%%%%%%%%%%%%%%%%%%%%%%%%%%%%%%%%%%%%%%
  \begin{figure*}
        \centering
        \includegraphics[width=0.48\linewidth]{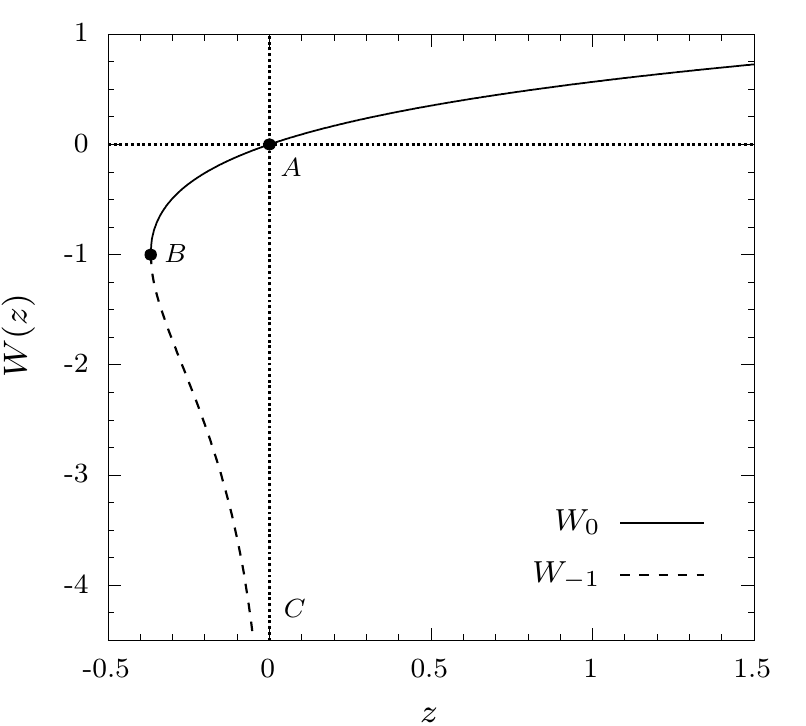}
        \quad\,\,\,
        \includegraphics[width=0.48\linewidth]{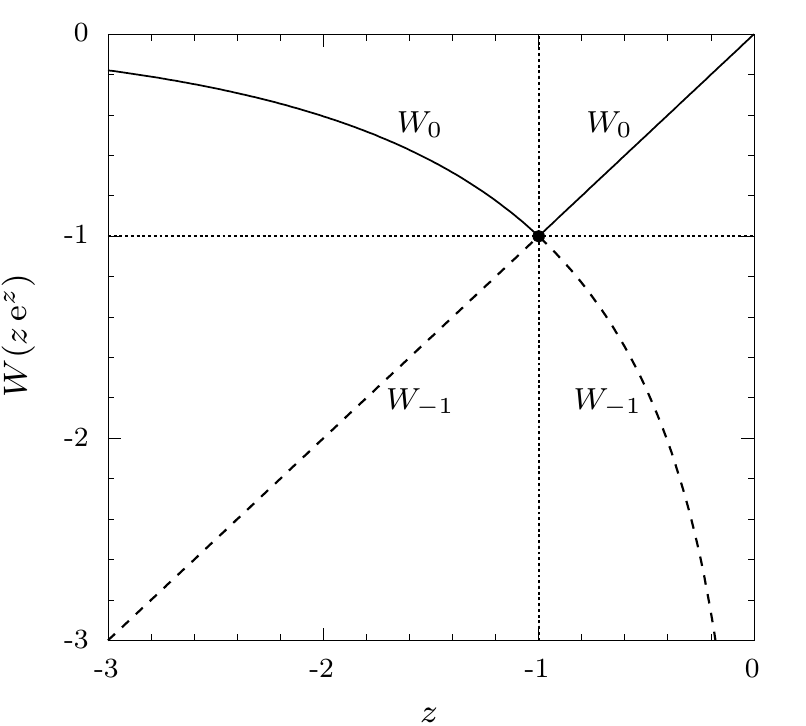}
        \caption{Left: the two real branches $\Wz$ (solid line) and $\Wmu$ (dashed line), 
                 where $A=(0\hspace{0.3mm},0)$ and $B=(-1/\e\hspace{0.2mm},-1)$, 
                 while C indicates the asymptotic point $(0\hspace{0.3mm},-\hspace{0.4mm}\infty)$.
                 Right: the two real branches of the function 
                 $W(\hspace{0.1mm}z\hspace{0.2mm}\e^z)$\hspace{0.2mm}.}
  \label{fig:Lambert}
  \end{figure*}
  %%%%%%%%%%%%%%%%%%%%%%%%%%%%%%%%%%%%%%%%%%%%%%%%%%%%%%%%%%%%%%%%%%%%%%%%%%%%%%

  The Lambert-Euler function is a multivalued function 
  defined implicitly by
  \begin{equation}
         W(z)\hspace{0.4mm}\e^{W(z)}=z\hspace{0.4mm},
         \quad\,\,\,
         z \in \mathbb{C}\hspace{0.25mm};
  \end{equation}
  the two real-valued branches
  of the $W$ are denoted as $\Wmu$ and $\Wz$ (see Fig. \ref{fig:Lambert}, left panel). 
  The asymptotic expansion of $\Wz$ reads 
  \begin{equation}
        \Wz(z)=
        \begin{cases}
              z + \Og(z^2)\hspace{0.4mm}, \hspace{1.8cm} z \to 0\hspace{0.4mm},
              \\[8pt]
              \ln z + \Og(\ln\ln z)\hspace{0.4mm}, \hspace{10.5mm} z \to \infty\hspace{0.2mm},
        \end{cases}
  \label{eq:Wz_asymp}
  \end{equation}
  (see e.g. de Bruijn 1981), while for $z \to 0$ it can be shown that 
  $\Wmu(z) \sim \ln\hspace{0.2mm}(-\hspace{0.4mm}z)$. Moreover, it can be proved that 
  \begin{equation}
        \begin{cases}
              \Wmu\hspace{-0.3mm}\left(z\hspace{0.4mm}\e^{\hspace{0.2mm}z}\right) = z\hspace{0.3mm},
              \quad\,\,
              \Wz\hspace{-0.2mm}\left(z\hspace{0.4mm}\e^{\hspace{0.2mm}z}\right) \geq z\hspace{0.3mm},
              \quad\,\, {\rm for} \,\,\,\, z \leq -\hspace{0.4mm}1,
              \\[8pt]
              \Wmu\hspace{-0.3mm}\left(z\hspace{0.4mm}\e^{\hspace{0.2mm}z}\right) \leq z\hspace{0.3mm},
              \quad\,\,
              \Wz\hspace{-0.2mm}\left(z\hspace{0.4mm}\e^{\hspace{0.2mm}z}\right) = z\hspace{0.3mm}, 
              \quad\,\, {\rm for} \,\,\,\, z \geq -\hspace{0.4mm}1.
        \end{cases}
  \label{eq:monotonicity}
  \end{equation}
  Therefore, $\Wmu\hspace{-0.2mm}\left(z\hspace{0.4mm}\e^{\hspace{0.2mm}z}\right) \leq z$, and
  $\Wz\hspace{-0.2mm}\left(z\hspace{0.4mm}\e^{\hspace{0.2mm}z}\right) \geq z$ for all values of $z$.
  Finally, we recall the monotonicity properties 
  $\Wz(z_1)\geq\Wz(z_2)$ and $\Wmu(z_1)\leq\Wmu(z_2)$ for $z_1 \geq z_2$. 
  For a general discussion of the properties of W, see e.g. Corless et al. (1996).
  
  In physics the $W$-\hspace{0.4mm}function has 
  been used to solve problems ranging from Quantum Mechanics (see e.g. Valluri et al. 2009;
  Wang \& Moniz 2019) to General Relativity (see e.g. Mez\H{o} \& Keady 2016; see also
  Barry et al. 2000 for a summary of recent applications), including Stellar Dynamics
  (CZ18). Indeed, several trascendental equations accuring in applications 
  can be solved in terms of $W$; for example, it is a simple exercise to prove that,
  for $X>0$, the equation
  \begin{equation}
        aX^b+c\ln X=Y\hspace{0.2mm},
  \label{eq:Lambert_gen}
  \end{equation}
  where $a$, $b$, $c$, and 
  $Y$ are quantities independent of $X$, has the general solution
  \begin{equation}
         X^b=\frac{c}{ab}\,W\!\left(\frac{ab}{c}\,\e^{\frac{b}{c}Y}\right)\hspace{-0.3mm}.
  \label{eq:W_eqApp}
  \end{equation}
  In particular, the solution of equation \eqref{eq:df/dx_iso} can be obtained
  for 
  \begin{equation}
        X=1+\frac{\xmin}{\xi},
        \quad\,\,
        a=b=1,
        \quad\,\,
        c=-\hspace{0.4mm}\frac{\MR}{2\xi},
        \quad\,\,
        Y=1+\frac{\chiup}{2\xi},
  \end{equation}
  as
  \begin{equation}
  1+\frac{\xmin}{\xi}=c\hspace{0.4mm}W\hspace{-0.4mm}\left(\frac{1}{c}\hspace{0.4mm}\e^{\frac{Y}{c}}\right)\hspace{-0.4mm}.
  \label{eq:z&m}
  \end{equation}
  We note that the equations \eqref{eq:xminISO} and \eqref{eq:z&m} represent the only
  solution for $\xmin$ in the isothermal accretion for generic values of the model
  parameters. This can be proved as follows. 
  The first condition for the general validity of \eqref{eq:z&m}
  is that the argument of $W$ must be $\leq 0$. In fact, $c \leq 0$ and
  $\xmin \geq 0$, so that the right hand side of \eqref{eq:z&m} must be 
  $\geq 0$, i.e. necessarily $W \leq 0$; from Fig. \ref{fig:Lambert} (left panel) this forces
  the argument to be $\leq 0$. This first condition is always true for our models. 
  The second condition, again from the left panel of Fig. \ref{fig:Lambert}, 
  is that the argument must be $\geq -\hspace{0.2mm}1/\e$ for all possible choices 
  of the model parameters. This inequality is easily verified by showing, with a standard
  minimisation of a function of two variables, that the minimum of the argument 
  over the region $Y \geq 1$ and $c \leq 0$ is indeed not smaller than $-\hspace{0.2mm}1/\e$. 
  Finally, we show that only the $\Wmu$ function appears in the solution for $\xmin$.
  This conclusion derives from the physical request that $\xmin \geq 0$, i.e. that
  the right hand side of equation \eqref{eq:z&m} is $\geq 1$.
  Let $z=1/c$. From the monotonicity properties of $\Wmu$ and $\Wz$ mentioned
  after equation \eqref{eq:monotonicity}, as $Y \geq 1$ we have 
  $z\hspace{0.4mm}\e^{\hspace{0.2mm}Yz} \geq z\hspace{0.4mm}\e^{\hspace{0.2mm}z}$, and 
  so equation \eqref{eq:monotonicity} yields 
  $\Wz\hspace{0.3mm}(z\hspace{0.4mm}\e^{\hspace{0.2mm}Yz}\hspace{0.1mm})
  \geq \Wz\hspace{-0.2mm}\left(z\hspace{0.4mm}\e^{\hspace{0.2mm}z}\right)
  \geq z$, i.e. 
  $c\hspace{0.2mm}\Wz\hspace{0.3mm}(\hspace{0.1mm}\e^{Y/c}\hspace{-0.3mm}/c) \leq 1$,
  being $z \leq 0$. An identical argument shows instead that 
  $c\hspace{0.2mm}\Wmu\hspace{0.2mm}(\hspace{0.1mm}\e^{Y/c}\hspace{-0.3mm}/c) \geq 1$,
  as required.
\end{document}